\documentclass[graybox]{svmult}

\usepackage{natbib}

\usepackage{mathptmx}
\usepackage{helvet}
\usepackage{courier}
\usepackage{type1cm}
\usepackage{makeidx}
\usepackage{graphicx}
\usepackage{multicol}
\usepackage[bottom]{footmisc}

\newcommand{\U}[1]{\ensuremath{\mathrm{~#1}}}

\newcommand{\yr}{\U{yr}}
\newcommand{\Myr}{\U{Myr}}
\newcommand{\Gyr}{\U{Gyr}}

\newcommand{\kpc}{\U{kpc}}

\newcommand{\hi}{H{\sc i} }

\newcommand{\HH}{\mbox{H$_2$}}
\newcommand{\Mo}{\mbox{M$_{\odot}$}}
\def\sbr{${\rm mag\,\,arcsec^{-2 }}$\ }

\newcommand{\Ha}{\mbox{H$_{\alpha}$}}
\newcommand{\cmm}{\mbox{cm$^{-2}$}}

\usepackage{ifthen}
\def\draftversion{1} 

\ifthenelse{\equal{\draftversion}{1}}{
	\usepackage{xcolor}
	\newcommand{\tmp}{}
	\newenvironment{envcomm}[1]{\renewcommand{\tmp}{#1}\begin{color}{blue}\begin{center}\hrule\vspace{0.5mm}\tmp's COMMENTS\end{center}}{\begin{center}END OF \tmp's COMMENTS\vspace{0.5mm}\hrule\end{center}\end{color}}
	\newenvironment{draft}{\begin{color}[rgb]{0,0.4,0}\begin{center}\hrule\vspace{0.5mm}DRAFT\end{center}}{\begin{center}END OF DRAFT\vspace{0.5mm}\hrule\end{center}\end{color}}
	\newcommand{\comcomm}[2]{\begin{color}{blue}\ $\bullet$ \textbf{#1:} #2 $\bullet$\ \end{color}}
	\newcommand{\revend}[1]{\par\begin{color}[rgb]{0,0.4,0}\begin{center}\hrule\vspace{0.5mm}END OF #1's REVISIONS\vspace{0.5mm}\hrule\end{center}\end{color}\par}
	\newcommand{\todo}[1]{\begin{color}{red}\ $\bullet$ \textbf{To do: }#1 $\bullet$\ \end{color}}
	
	}{
	\newsavebox{\trashcan}
	\newenvironment{envcomm}[1]{\begin{lrbox}{\trashcan}\begin{minipage}{\columnwidth}}{\end{minipage}\end{lrbox}}
	
	\newcommand{\comcomm}[2]{}
	\newcommand{\revend}[1]{}
	\newcommand{\todo}[1]{}
	
	}

\makeindex

\begin{document}

\title*{Tides in colliding galaxies}
\author{Pierre-Alain Duc \and Florent Renaud}
\institute{Pierre-Alain Duc \at AIM Paris-Saclay, CNRS/INSU, CEA/Irfu, Universit\'e Paris-Diderot, Service d'astrophysique, Orme des Merisiers, 91191 Gif sur Yvette cedex, France;  \email{paduc@cea.fr}
\and Florent Renaud \at Observatoire de Strasbourg, CNRS UMR 7550 and AIM Paris-Saclay, CEA/Irfu, CNRS/INSU, Universit\'e Paris-Diderot, Service d'astrophysique, Orme des Merisiers, 91191 Gif sur Yvette cedex, France;  \email{florent.renaud@cea.fr}
}

\maketitle


\abstract*{}

\abstract{Long tails and streams of stars are the most  noticeable upshots of galaxy collisions. Their origin as gravitational, tidal, disturbances has however been recognized only less than fifty years ago and more than ten years after their first observations.  This Review describes how the idea of galactic tides emerged, in particular thanks to the advances in  numerical simulations, from the first ones that included tens of particles to the most sophisticated  ones with tens of millions of them and state-of-the-art hydrodynamical prescriptions.  Theoretical aspects pertaining to the formation of tidal tails are then presented. The third part of the review turns to observations and underlines the need for collecting deep multi-wavelength data to tackle the variety of physical processes exhibited by collisional debris. Tidal tails are not just stellar structures, but turn out to contain all the components usually found in galactic disks, in particular atomic / molecular gas and dust. They host star-forming complexes and are able to form star-clusters or even second-generation dwarf galaxies.
The final part of the review discusses what tidal tails can tell us (or not) about the structure and content of present-day galaxies, including their dark components,  and explains how tidal tails may be used to probe the past evolution of galaxies and their mass assembly history. On-going deep wide-field surveys disclose many new low-surface brightness structures in the nearby Universe, offering great opportunities for attempting galactic archeology with tidal tails.}

\section{Preliminary remarks}

The importance of tides on bodies  in the Solar System  has been understood and quantified  for many decades. The various contributions in this Volume reflect the maturity of this field of research. Advances in the appreciation of the role of tidal effects on planet/stellar evolution are also remarkable. As the extragalactic world is concerned, the situation is paradoxical. Whereas the effects of tidal forces are spectacular -- they alter the morphology of the most massive galaxies and may lead to the total destruction of the dwarf satellite galaxies -- , it is only in the seventies that tides were recognized as actors of galactic evolution. Observations of jet-like structures, antennas, bridges and plumes occurred well before they were interpreted as ``tidal tails". Only the first numerical simulations of galaxy mergers convinced the community about the real nature of these stellar structures, whereas the straightforward consideration that galaxies are flaccid bodies might have lead to this conclusion much earlier. It is however true that the bulges generated by the Moon and the Sun on the Earth's oceans, which were interpreted as the result of tides soon after the laws of gravity were established,  do not resemble the gigantic appendices that emanate from some galaxies although their origin is similar. What seems obvious now was not fifty years ago.

Having said that, it would be misleading to claim that tidal forces are the only  actors of  galactic morphological transformations. In fact, the fraction of mass involved in material that is tidally affected is relatively small. Other physical processes such as violent relaxation are more important in shaping galaxies. The nuclear starbursts often associated with galaxy mergers are not directly induced by tidal forces. Furthermore, not all the collisional debris found around mergers are, strictly speaking, of tidal origin. 
With these preliminary remarks, we wish to precise that this Review specifically focusses on tides in colliding galaxies and is not an overview of interacting galaxies and associated phenomena. For a more general insight on galaxy-galaxy collisions and mergers, the reader is referred to the somehow old but comprehensive reviews of \cite{Sanders96b} and \cite{Schweizer98}, dealing with observations, and \cite{Struck1999}, more focussed on simulations.

We will first present the historical context of the discovery of tails around galaxies, and detail how the role of tides  became evident. The tremendous progress in the numerical modeling of tidal tails is detailed before a more theoretical and analytical approach of the formation of tidal tails is presented.
In the following sections, we investigate the physical properties of tidal tails, emphasizing what deep multi-wavelength observations bring to their study. 
We then make close up on the tails, looking at their sub-structures: from young stars and star clusters to tidal dwarf galaxies.
Finally, we examine what tidal features may tell us about galaxies: what they are made of, and how and when they were formed. 
We hope to convince the reader that tails are not only aesthetic add-ups in images of colliding galaxies but may be used to address  fundamental questions of astrophysics.

\section{Historical context}

In the late 1920s, the observational power of 100-inches (2.5 m) class telescopes allowed Hubble to determine the existence of apparently isolated nebulae outside of our Milky-Way \citep{Hubble1929}. These so-called ``island universes'' became of prime importance in the discovery of the expansion of the Universe, thanks to redshift measurements. Rapidly, many more extra-galactic objects have been classified as galaxies and sorted according to their morphology, following the famous Hubble pitch fork diagram.

\subsection{Discovery of peculiarities}
\label{sec:pec}

In the preface of his Atlas of Peculiar Galaxies, \citet{Arp1966} noted that ``when looked at closely enough, every galaxy is peculiar''. While most of the luminous galaxies could be classified as either elliptical, spiral or barred-spiral, it appeared that more and more peculiar morphologies would not fit into these three families. Number of photographic plates of individual systems have been published and revealed twisted shapes and/or faint extensions outside of the central regions of the galaxies \citep[e.g.][]{Duncan1923, Keenan1935, Wild1953, Zwicky1956}. These features have been detected in many other objets gathered in atlases and catalogues \citep{Zwicky1961, ZwickyHerzogWild1963, VorontsovVelyaminovArkhipova1962, VorontsovVelYaminov1964, Arp1966}. This contradicted the persistent idea that the intergalactic space was entirely empty (\citealt{Zwicky1963}; see also the discussion in \citealt{Gold1949} and references therein).

It rapidly appeared that many of these peculiar galaxies were actually double or multiple galaxies, i.e. pairs or small groups, observed close to each other. Really interacting galaxies have been told apart from optical pairs, for which apparent closeness is due to projection effects (\citealt{Holmberg1937}; see also \citealt{Zwicky1956}). The major signatures of interaction were the detection of long ($\sim 10^{1-2} \kpc$) and thin ($\sim 1 \kpc$) filaments either connecting two galaxies or pointing away from them. The former have been named bridges and the latter, tails. This clearly distinguished them from the spiral arms which are located in the more central regions of disk galaxies. However, a confusion persisted because it was noted that tails are sometimes (but not always) in the continuation of spiral arms \citep{PikelNer1965}. Although being faint and thus often difficult to observe, these filaments appear bluer than the disks themselves, suggesting that they host ongoing star formation \citep{Ambartsumian1961, Zwicky1963}. But the exact reasons for such morphological features remained opened to debate over the entire 1960 decade.

\subsection{A controversial scenario}
\label{sec:controv}

\citet{Zwicky1962} proposed that collisions of galaxies would enhance the supernovae activity, by increasing the probability of chain explosions. These blasts could then sweep out or eject the galactic material away from the nuclei. With a favorable geometry, such events could even act as ``launchers of galaxies'', and thus account for the intergalactic filamentary structures. However, this scenario failed to explain the thinness of the filaments and the connection to other galaxies, so that it has rapidly been ruled out \citep{PikelNer1965}.

\begin{figure}
\includegraphics[width=\textwidth]{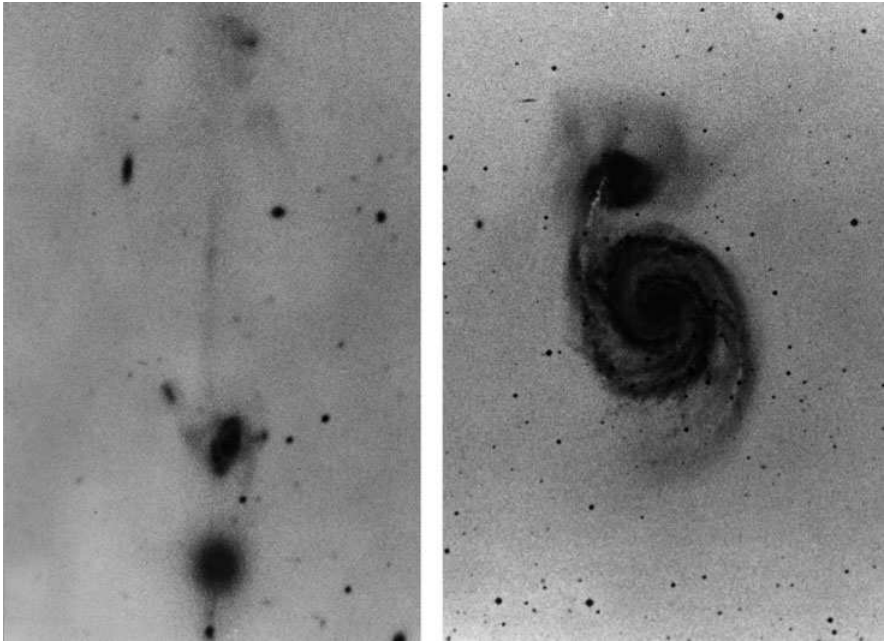}
\caption{NGC~3561 (left) has been seen by \citet{Arp1972} as a spiral galaxy having ejected two luminous jets of matter. An high surface brightness object, called ``Ambartsumian's knot'', can be seen at the tip of the southern jet at the bottom-edge of this image. In the case of M~51 (right), the companion is situated at the tip of a spiral arm of the main galaxy. Images from the Atlas of Peculiar Galaxies by H. Arp, available in the NASA/IPAC Extragalactic Database, Level 5.}
\label{fig:jets}
\end{figure}

Another explanation for the formation of bridges took jets into account \citep{Ambartsumian1961, Arp1967, Arp1968, Arp1972}. When a massive galaxy ejects a fraction of its matter (gaseous, stellar or both) from its nucleus, a symmetrical pair of jets is formed but rapidly slowed down by the high densities encountered along its path\footnote{According to \citet{Arp1969}, the same mechanism would account for the creation of spiral arms in rotating galaxies.}. This would create an overdensity at the tip of the jets that could condense and form a small companion galaxy \citep{Sersic1968}. All together, the main galaxy, its companion and one of the jets would constitute the interacting pair and the bridge. The absence of galaxy at the end of the second jet (i.e. the tail) was explained by either the escape of the companion to the intergalactic medium, its rapid dissolution, or a delayed formation that has not taken place yet \citep{Arp1969}. Illustrative examples of this scenario are NGC~3561 (``the guitar'') and M~51 (``Whirlpool galaxy''), as shown in Figure~\ref{fig:jets}. However, \citet{Holmberg1969} noted that the condensation of the gravitationally bound companion galaxy would be very unlikely when the jets reach a velocity higher than the escape velocity, which seems to be true in most of the cases. Such an activity from the nuclei of massive galaxies led some authors to classify galaxies with connecting ``jets'' as radio-galaxies \citep[see e.g.][]{Ambartsumyan1974}.

Meanwhile, tides have been considered as a possible cause for the filaments: a close passage of one galaxy next to another would lead to different gravitational forces over the spatially extended galaxies: the side of the first galaxy facing the second is more attracted than the opposite side. These differential forces would then significantly deform the shape of the galaxies and could even trigger an exchange of some of their stars \citep{Holmberg1941, Zwicky1953, Zwicky1956, Lindblad1960, Zasov1968, Tashpulatov1970a}. The pioneer numerical works that addressed this question concluded that, under precise circumstances, tidal structures looking like bridges and tails could form during the close encounter of two galaxies \citep{Lindblad1961, Yabushita1971}.

However, the tidal origin of the tails has been intensively discussed. \citet{VorontsovVelYaminov1962} argued that the elongation of the tails (sometimes up to a few $\times 100 \kpc$, see \citealt{Mirabel1991}) was too large to be produced by tides. He added that close pairs of galaxies were not systematically linked to the existence of filaments, and concluded that tails and bridges shared the same origin than the more classical spiral arms. Others followed the same line of arguments and evoked magnetic (or magnetic-like, see \citealt{VorontsovVelYaminov1965}) fields to explain the narrow shape of the tails \citep[see e.g.][]{Burbidge1963, Zasov1968}. Tubes of magnetic lines forming at the same time as the galaxy itself would propagate a wave that would trigger the condensation of gas along them. Such an hypothesis would explain the presence of knots of high surface brightness along the tails, as already detected by e.g. \citet{Burbidge1959}. Furthermore, \citet{Gershberg1965} noted that a collision between two galaxies would heat up the gas too much ($\sim 10^7 \U{K}$) to form a thin structure and ruled out this scenario as a possible cause of creation of filaments. \citet{Arp1966} summarized the debate by suggesting that forces other than pure gravitation should be at stake in the shaping of peculiar galaxies and their intergalactic structures.

\subsection{Tidal origin}

The major breakthrough came in the early 1970s, in the newly-born era of computers. Thanks to a series of numerical experiments, \citet{Toomre1972} showed that the brief but intense tidal forces arising during the encounter of two disk galaxies would be sufficient to create structures as long and thin as the tails referenced in the catalogues. They extended the works of \citet{Pfleiderer1963} and \citet{Tashpulatov1970a, Tashpulatov1970b} by considering a bound companion galaxy on an very eccentric orbit, as well as disks inclined with respect to the orbital plane. In their study, a single galaxy is represented by a point-mass surrounded by rings of test particles whose masses are zero. When two of such galaxies are set on a given orbit, the central point-mass follows Kepler's law of motion. The test particles feel the net gravitational potential and thus, their motion is affected by both point-masses. However, in this method called restricted simulation, the mass-less test particles themselves do not affect the gravitational field of the galaxy. 

\begin{figure}
\includegraphics[width=\textwidth]{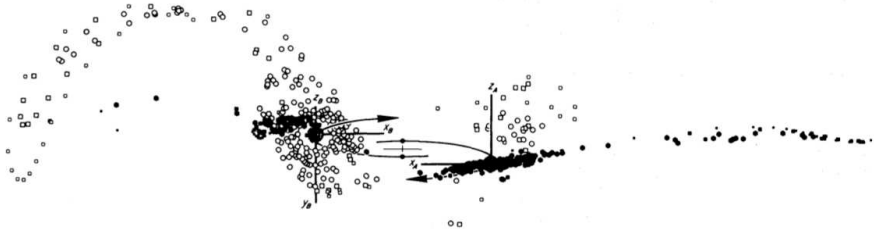}
\caption{Restricted simulation of the Mice galaxies (NGC~4676) from \citet{Toomre1972}. The two tails that exhibit very different shapes and thickness have been successfully reproduced numerically by considering tidal interaction only. Images of the real galaxies are shown in Figure~\ref{fig:toomre_sequence}, second panel, and  Figure~\ref{fig:HI}, panel 4.}
\label{fig:mice_toomre}
\end{figure}

\citet{Toomre1972} noted that close passages could induce a deformation of the disk(s), possibly leading to the creation of bridges and/or tails. By varying several parameters of the problem such as the inclination of the disks or the eccentricity of the orbit, they have shown that gravitation only was enough to reproduce the structures observed in interacting systems (see Figure~\ref{fig:mice_toomre}). This showed the way to many other numerical experiments \citep{Eneev1973, Lauberts1974, Keenan1975} and allowed to conclude on the tidal origin of several observed features \citep{Danziger1974, Stockton1974, Yabushita1977}.

Since then, gravitational tides have been considered as the major cause of the creation of filaments in interacting galaxies. That is why such features are often refered to as tidal structures. 

\begin{figure}
\includegraphics[width=\textwidth]{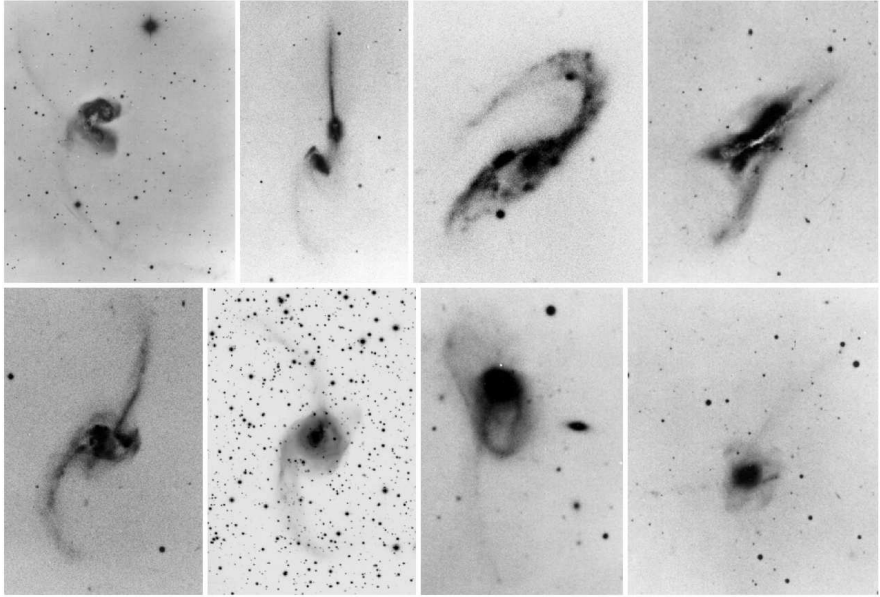}
\caption{The Tommre's sequence represents the supposed evolution of interacting galaxies. It starts with the early phases, when progenitors have just begun to interact, shows intermediate stages and finishes with the coalescence phase. From left to right, top: NGC~4038/39 (the Antennae), NGC~4676 (the Mice), NGC~3509, NGC~520; bottom: NGC~2623, NGC~3256, NGC~3921, NGC~7252 (the Atoms for Peace). Images from the Atlas of Peculiar Galaxies by H. Arp, available in the NASA/IPAC Extragalactic Database, Level 5. }
\label{fig:toomre_sequence}
\end{figure}

An examination of the peculiar galaxies with the new light shed by numerical experiments on tides revealed that most of these galaxies would fit into an evolutionary sequence (see Figure~\ref{fig:toomre_sequence}), called Toomre's sequence \citep{Toomre1977}. Each step represents a dynamical stage in the evolution of interacting galaxies toward the final coalescence of the merger\footnote{Note that the position of some of the galaxies in the sequence has been recently discussed thanks to new numerical models \citep[see e.g.][]{Karl2010}.}. With time going, the tidal features created by the first encounters slowly vanish into the intergalactic medium or are captured back by their galaxy. Note however that relics of the tails remain visible for several $10^9 \yr$ \citep{Springel1999}.

\subsection{Forty years of numerical simulations}

In order to retrieve the steps of the Toomre's sequence and to better understand the role of each parameter involved in interacting galaxies, an important amount of work has been conducted by many authors since the very first (non-numerical) computations in the early 40's. At that time, \citet{Holmberg1941} used the light and the property of the decay of its intensity as $r^{-2}$ as a proxy for gravitation. He set a pair of two ``nebulae'', each made of 37 light-bulbs, and computed the equivalent gravitational acceleration by measuring the intensity of the light thanks to galvanometers at several positions. This ingenious method allowed him to spot the creation of ``spirals'' during a close encounter. But it is in the numerical era that most of the progresses have been done.

Despite their success in reproducing observed systems, the restricted simulations of \citet{Toomre1972} lacked the orbital decay due to dynamical friction. The problem was solved when considering self-consistent (``live'') galaxies, i.e. models where all the particles interact with each other \citep[e.g.][]{White1978, Gerhard1981}. However, the cost of such computations was very high at that time. That is why tree-codes \citep{Barnes1986, Hernquist1987} and multipole expansions techniques \citep{vanAlbada1982, White1983} have been introduced to decrease the computation time, or equivalently to increase the reachable  resolution.

\citet{Barnes1988} presented the first simulation of self-consistent multi-components galaxies. He showed that the presence of a dark-matter halo increases significantly the dynamical friction, thus favoring the merger of the galaxies.

In the same time, \citet{HernquistKatz1989} gathered the tree-code method and the smooth particle hydrodynamics (SPH) technique \citep{Lucy1977, Gingold1977} to treat both the gravitation and the hydrodynamics within a particle-based code. In SPH simulations, the physical properties of the particles are smoothed over a kernel of finite size, centered on the particle itself. Thanks to this Lagrangian approach, SPH does not suffer from the limitations of grid codes \citep{Hockney1988}, i.e. mainly the waste of computational power in areas of nearly vacuum, an omnipresent situation in the case of galaxy mergers. In a similar way, the so-called ``sticky-particle'' method considers clouds as collisionless particles. When two clouds are in a close encounter, they loose energy via dissipation, mimicking an inelastic collision \citep{Negroponte1983}.

Following this idea, \citet{Noguchi1986} proposed a galaxy model made of two types of particles: gaseous clouds and stars. When such a galaxy interacts with a point-mass encounter, these authors found the cloud-cloud collisions to be more frequent, and considered this as a burst of star formation (mostly at the times of the pericenter passages of the progenitor galaxies). \citet{Mihos1991, Mihos1992, Mihos1993} took one step further by considering the interstellar media (ISM) of both galaxies and monitored their interaction to characterize the formation of stars. They took advantage of the dissipative nature of their models to show that the merger phase could take place up to twice faster than in gas-free simulations.

Since then, a lot of flavors of these methods has been widely applied to many topics. Some improvements also appeared, to speed up the computation and thus to allow higher resolutions \citep[see e.g.][]{Dehnen2000}. More and more hybrid codes take advantage of multiples methods \citep[e.g.][]{Semelin2002, Berczik2003} to increase accuracy and speed-up.

\begin{figure}
\includegraphics[width=\textwidth]{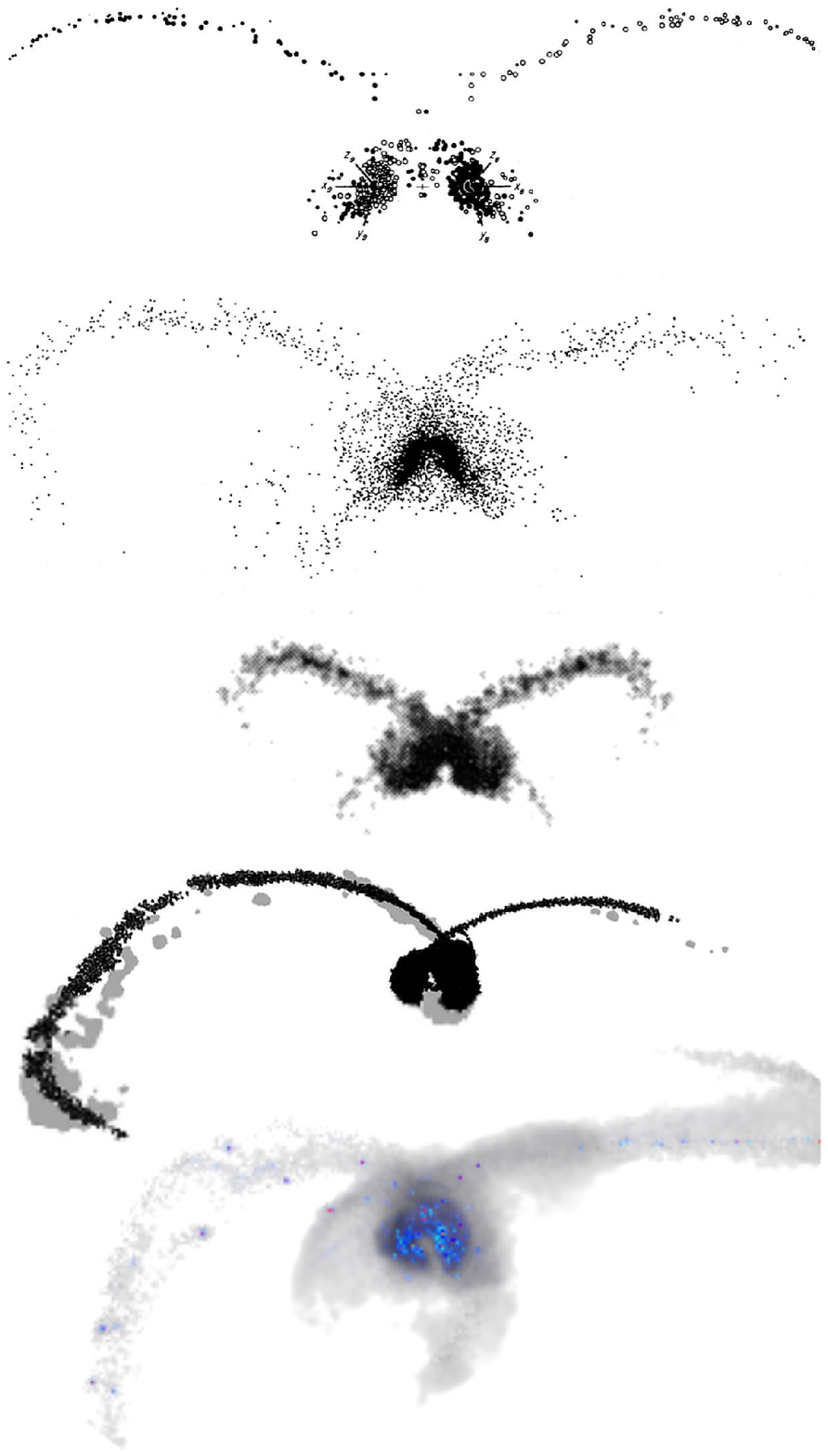}
\caption{Numerical simulations of the Antennae galaxies (NGC~4038/39) within four decades. From top to bottom: restricted simulation of \citet{Toomre1972}; first self-consistent simulation of the Antennae by \citet{Barnes1988}; hydrodynamic run of \citet{Mihos1993}; recent models with SPH by \citet{Karl2010} and with AMR by \citet{Teyssier2010}. Improvements in both the techniques and the set of parameters allowed the models to get closer and closer to the observational data (see Figure~\ref{fig:HI}, panel 6).}
\label{fig:antennae_runs}
\end{figure}

Recently, the adaptive mesh refinement (AMR) technique has been used for modeling a merger of two gas-rich galaxies at high resolution \citep{Kim2009, Teyssier2010}. AMR codes combine the power of the Lagrangian approach where dense regions are highly resolved, and the continuous description of the ISM on grids \citep[e.g.][]{Fryxell2000, Teyssier2002, Oshea2004}. The computational domain is meshed on a (usually catesian) grid, which is refined at the regions of interest, typically those of highest densities. Two different snapshots of a numerical model using the AMR technique are shown on Figure~\ref{fig:antennae_runs}  and Figure~\ref{fig:sim-sscs} (left panel). 

As seen in the literature since \citet{Toomre1972}, simulations of interacting galaxies can follow two approaches:
\begin{itemize}
\item the systematic exploration of a large volume of the parameter space, with the goal of understanding the influence of certain parameters on the evolution of the merger and its stellar population \citep[see e.g.][]{Olson1990, Wallin1992, Springel2000, Naab2003, Gonzalez2005, Kapferer2005}. Among them, the GalMer project \citep{DiMatteo2007} gathers $\sim 1000$ SPH simulations of mergers and the associated star formation histories, and makes them publicly available online\footnote{http://galmer.obspm.fr/}. With such databases, the simulations can be interpreted statistically, thus strengthening the physical conclusions.

\item the simulation of specific, observed galaxies in order to bring new lights when interpreting the observations \citep[see e.g.][]{Barnes1988, Mihos1993, Hibbard1995, Duc2000, Barnes2004, Bekki2008, Renaud2008, Dobbs2010, Karl2010, Teyssier2010}. Several pairs of interacting galaxies have been numerically reproduced (see an example in Figure~\ref{fig:antennae_runs}) by putting the effort on finding a set of parameters that best describe the pair, generally by trial-and-error. Intuition and experience play an important role in such a study. However, this process has recently been automatized thanks to new numerical tools: these codes make a series of restricted, fast simulations (``{\`a} la Toomre \& Toomre'') and slighly modify one parameter of the initial conditions at each iteration, to improve the match with observational data (see for example Identikit by \citealt{Barnes2009}, Barnes 2011, in prep., and AGC by \citealt{Smith2010}). Genetic algorithms have also been implemented to optimized the search of the parameters \citep[see the MINGA code by][]{Theis2001, Theis2003}. This way, a large range of parameters can be covered very quickly to find which set best matches the observational data. However because the simulations are restricted, they do not account for the orbital decay of the galaxies due to dynamical friction, which represents an important limitation for such methods, in particular when multiple collisions occur. Starting from the set of parameters suggested by such fast codes and fine-tuning them in self-consistent re-runs could be a good compromise.
\end{itemize}

The simulation of interacting galaxies is not limited to pairs. However, numerical models of observed (compact) groups of galaxies are still very rare, due to the difficulty to set a consistent scenario for an entire group. Each galaxy-galaxy interaction has to take place in a system already perturbed by the previous interactions, such that the mass and the orbit of the progenitor are to be re-evaluated constantly during the evolution of the group. (Some attempts have been made in the case of Stephan's Quintet, see \citealt{Renaud2010, Hwang2011}).

While they face an increasing need of resolution and accuracy, these state-of-the-art numerical methods can efficiently provide a solution to the questions raised by observations at higher and higher resolution. Simulations of interacting galaxies still represent an important part of the numerical work done in astrophysics. The models of individual galaxies are regularly updated to fit the most recent theories on galaxy formation and evolution and include better descriptions of the physical processes. Nowadays, the research on interacting galaxies is mainly threefold:
\begin{itemize}
\item The cosmological approach, mostly based on the $\Lambda$ Cold Dark Matter (CDM)  theory, focusses on galaxy formation through repeated accretion of satellites \citep[e.g.][]{Neistein2008, Cole2008}. In particular, this hierarchical scenario describes the formation of elliptical galaxies as remnants of a merger but also provides clues on the dynamical status and evolution of groups of galaxies.
\item A growing number of works focus on the central region of mergers. The formation of active galaxy nucleus (AGN) and the associated feedback is intensively discussed, as well as the pairing of black holes in mergers \citep[see e.g.][among many others]{Lagos2008, Narayanan2008, Blecha2011, Debuhr2011, Sijacki2011}. 
\item The stellar populations of the interacting galaxies and the properties of the star clusters and the dwarf galaxies they may contain is also a widely covered topic \citep[e.g.][]{Wetzstein2007, Bournaud2008, Dobbs2010}. In this respect, the tides play an important role on the physics of these subsystems. This last point will be further developed in the following sections.
\end{itemize}

\section{Theory of the tidal tail formation in interacting galaxies}

After having reproduced numerically some of the extragalactic tidal structures observed in the Universe, several physical and mathematical descriptions of the phenomenon have been proposed to better understand the tides at galactic scale. The complexity of the task comes from the diversity of possible configurations, which translates into a large number of parameters. In this section, we review the role of the first order parameters and illustrate their respective effects thanks to numerical simulations of interacting galaxies. A mathematical description of the tidal field is also presented.

\subsection{Gravitational potential and tidal tensor}

By definition, the tides are a differential effect of the gravitation. Let's consider a galaxy, immersed in a given gravitational field. At the position of a point within the galaxy, the net acceleration can be split into the effect from the rest of the galaxy $\vec{a}_\mathrm{int}$, and the acceleration due to external sources $\vec{a}_\mathrm{ext}$. The latter can itself be seen as a part common to the entire galaxy (usually the acceleration of the center of mass), and the differencial acceleration, that differs from point to point within the galaxy. In other terms, the net acceleration at the position $r_\mathrm{P}$, in the reference frame of the center of mass of the galaxy (which lies at the position $r_\mathrm{g}$), is given by
\begin{equation}
\vec{a}(\vec{r}_\mathrm{P}) = \vec{a}_\mathrm{int}(\vec{r}_\mathrm{P}) + \left[ \vec{a}_\mathrm{ext}(\vec{r}_\mathrm{P}) - \vec{a}_\mathrm{ext}(\vec{r}_\mathrm{g}) \right].
\end{equation}
For small $\vec{\delta} = \vec{r}_\mathrm{P} - \vec{r}_\mathrm{g}$ with respect to $\vec{r}_\mathrm{g}$, one can develop at first order and get
\begin{equation}
\vec{a}(\vec{r}_\mathrm{P}) = \vec{a}_\mathrm{int}(\vec{r}_\mathrm{P}) + \vec{\delta}\ \D \vec{a}_\mathrm{ext},
\end{equation}
which also reads
\begin{equation}
\left(\begin{array}{l}
a_x(\vec{r}_\mathrm{P})\\
a_y(\vec{r}_\mathrm{P})\\
a_z(\vec{r}_\mathrm{P})
\end{array}\right)
= \left(\begin{array}{l}
a_{\mathrm{int}, x}(\vec{r}_\mathrm{P})\\
a_{\mathrm{int}, y}(\vec{r}_\mathrm{P})\\
a_{\mathrm{int}, z}(\vec{r}_\mathrm{P})
\end{array}\right)
+ \left(\begin{array}{l}
\left[\delta_x \frac{\D}{\D x} + \delta_y \frac{\D}{\D y} + \delta_z \frac{\D}{\D z} \right] a_{\mathrm{ext}, x}\\
\left[\delta_x \frac{\D}{\D x} + \delta_y \frac{\D}{\D y} + \delta_z \frac{\D}{\D z} \right] a_{\mathrm{ext}, y}\\
\left[\delta_x \frac{\D}{\D x} + \delta_y \frac{\D}{\D y} + \delta_z \frac{\D}{\D z} \right] a_{\mathrm{ext}, z}
\end{array}\right),
\end{equation}
or simpler
\begin{equation}
a^i_\mathrm{P} = a^i_\mathrm{int}(\vec{r}_\mathrm{P}) + \delta_j\ \partial^j a^i_\mathrm{ext},
\end{equation}
when using Einstein's summation convention. The effect of the external sources on the galaxy are described by the term
\begin{equation}
T^{ji} \equiv \partial^j a^i_\mathrm{ext},
\end{equation}
which is the $j,i$ term of the $3\times 3$ tensor $\tens{T}$ called tidal tensor \citep{Renaud2008}. Such a tensor encloses all the information about the differential acceleration within the galaxy. Therefore, the (linearized) tidal field at a given point in space is described by the tensor evaluated at this point. 

Note that the tidal tensor is a static representation of the tidal field: the net effect on the galaxy also depends on its orbit in the external potential, or in other words, on the variations of intensity and orientation of the tidal field. This can be accounted for by writing to pseudo-accelerations (centrifugal, Coriolis and Euler) in the co-rotating (i.e. non-inertial) reference frame, or by the means of a time-dependent effective tidal tensor in the inertial reference frame. For simplicity, in the following we focus on static, purely gravitational tides and refer the reader to \citet{Renaud2011} for more details.

Because the acceleration $\vec{a}_\mathrm{ext}$ derives from a gravitational potential $\varphi_\mathrm{ext}$, one can write
\begin{equation}
T^{ji} = -\partial^j\partial^i \varphi_\mathrm{ext} = -\partial^i\partial^j \varphi_\mathrm{ext} = T^{ij}.
\label{eqn:tensor}
\end{equation}
(Several examples of tidal tensors of analytical density profiles are given in \citealt{Renaud2009}, see also the Appendix~B of \citealt{Renaud2010PhD}.) It is important to note that these considerations are scale-free and applies to any spatially extended object, such as galaxy clusters, galaxies, star clusters, stars, planets, etc.

\begin{figure}
\includegraphics[width=\textwidth]{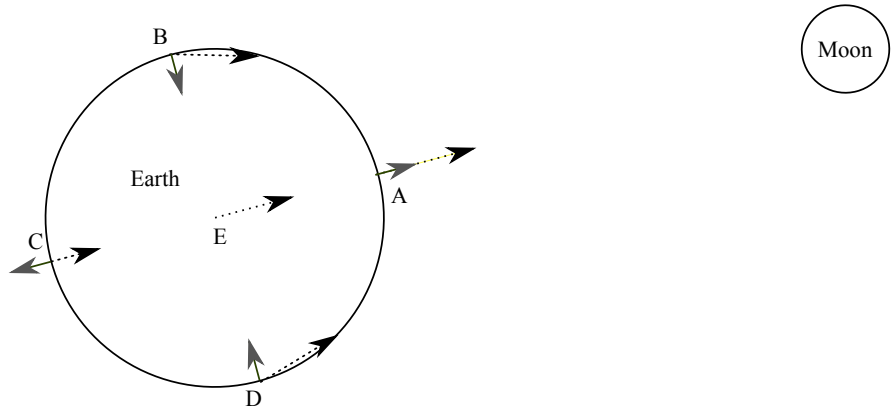}
\caption{Gravitational attraction (black, dotted line) of the Moon on the Earth, and the differential forces (grey). The tidal effect appears to be extensive in A and C, while it is compressive in B and D.}
\label{fig:earthmoon}
\end{figure}

For example, let's consider the Earth-Moon system and compute the tidal field with the Moon as source of gravitation. It can been seen from the Earth as a point-mass, and yields a potential of the form
\begin{equation}
\varphi_\mathrm{ext} = -\frac{GM}{r},
\end{equation}
with $r=\sqrt{x_i^2+x_j^2+x_k^2}$. The components of the tidal tensor are
\begin{equation}
T^{ij} = \frac{GM}{r^5} \left(3 x_i x_j - \delta^{ij} r^2\right)
\end{equation}
where $\delta^{ij} = 1$ if $i=j$ and 0 otherwise. When computed at the distance $d$ along the $i$-axis (i.e. for $r = d$ and $x_j = x_k =0$), the tidal tensor becomes
\begin{equation}
\tens{T}(d, 0,0) = \frac{GM}{d^3} \left(\begin{array}{rrr}2 & 0 & 0\\ 0 & -1 & 0\\ 0 & 0 & -1\end{array}\right).
\end{equation}
The signs of the diagonal terms (which are, in this case, the eigenvalues because the tensor is writen in its proper base) denotes differential forces pointing inward along the $i$-axis, and outward along the other two axes. A rapid study of the differential forces around the Earth (see Figure~\ref{fig:earthmoon}) shows indeed, that they point toward the Earth along the axes perpendicular to the direction of Moon. One speaks of a compressive effect. Along the Earth-Moon axis however, the differential forces point away from the planet: the effect is extensive.

\subsection{Compressive tides}

Back to the general case, it follows from Equation~\ref{eqn:tensor} that any tidal tensor is symmetric. Because it is also real-valued, it can be set in diagonal form, by switching to its proper base. In this case, three eigenvalues $\{\lambda_i\}$ denote the strength of the tides along the associated eigenvectors. The trace of the tensor (which is base-invariant) reads
\begin{equation}
\mathrm{Tr}(\tens{T}) = \sum_i \lambda_i = -\partial^i\partial^i \varphi_\mathrm{ext} = -\nabla^2 \varphi_\mathrm{ext}, 
\end{equation}
which can be connected to the local density $\rho$ thanks to Poisson's equation:
\begin{equation}
\mathrm{Tr}(\tens{T}) = -4\pi G \rho \le 0.
\end{equation}
The condition on the sign of the trace implies that it is impossible to compute simultaneously three strictly positive eigenvalues. Remains the cases of two, one or no positive eigenvalues, as mentioned by \citet{Dekel2003}. For two or one positive $\lambda$'s, the tidal field is called (partially) extensive, like e.g. in our Earth-Moon example. When all three eigenvalues are negative, the tides are (fully) compressive. By noticing that $\tens{T}$ is minus the Hessian matrix of the potential, one can show that a change of curvature of the potential implies a change of sign for $\tens{T}$. Therefore, compressive tides are located in the cored regions of potentials only, and never in cusps.

Note that a compressive mode (three negative $\lambda$'s) implies that the local density due to the source of gravitation is non-zero. Although such a situation does not exist with point-masses, it can occurs when considering extended mass distributions, like e.g. for galaxies embedded in a dark matter halo.

The duality of compressive/extensive tidal modes plays a role in the formation, early evolution and dissolution rates of star clusters. It has been noted that observed young clusters were preferentially found in the regions of compressive tides \citep[see][in the case of the Antennae galaxies]{Renaud2008}, and a compressive mode would slow down the dissolution of young globulars \citep{Renaud2011}.

\subsection{Formation of tidal tails and bridges}

In isolation, a galaxy keeps its material, which is made of dark matter, stars, gas and dust, bound thanks to the gravitation. However, when it moves in an external potential, created for instance by neighbor galaxies, it can experience gravitational forces which are different from one side of the galaxy to the other. In other words, the galaxy is plunged in a tidal field. As a result, its material undergoes deforming effects that re-arrange the individual components of the galaxy. On the one hand, when this material was initially distributed in an (almost) random way in phase-space (as opposed to e.g. sharing a common velocity pattern), the net tidal effect does not translate into a clear global change for an entire region of the galaxy. Therefore, such tides are difficult to detect. On the other hand, when large scale, regular patterns exists in the distribution of the galactic material in phase-space (e.g. a disk), the tides have a similar impact on stars that already lied in the same region of phase-space. All these stars are affected the same way and thus, the effect is much more visible. In the end, a given tidal field is easier to detect when it affects a regular, organized distribution of matter, than when it applies to isotropic structures. This is the reason why tidal features like tails and bridges are well visible around disk galaxies where the motion is well-organized, and merely inexistent in ellipticals, which yield much more isotropic distributions of positions and velocities. This last point can be extended to all structures with a high degree of symmetry (halos, bulges, and so on), as opposed to axisymmetric components like disks.

\begin{figure}
\includegraphics[width=\textwidth]{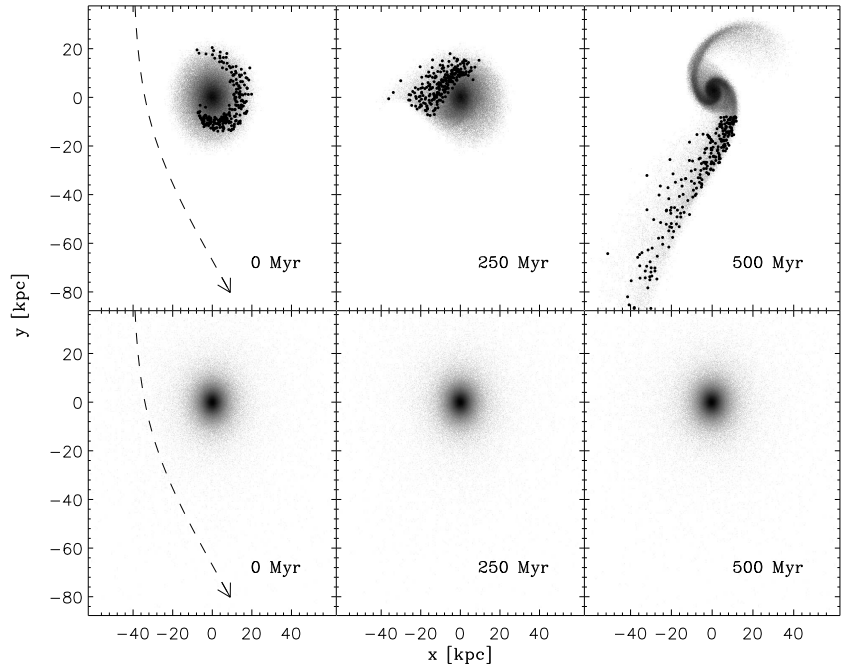}
\caption{Top: morphology of a disk galaxy, seen face-on, during its coplanar interaction with a point-mass (mass ratio = 1), before the interaction (left), at pericenter (middle) and after (right). The dashed line indicates the trajectory of the point-mass (from top to bottom). The black dots tag a subset of particles that are situated in one of the tidal tails at $t =500 \Myr$. Bottom: same but for an elliptical galaxy. No tidal structures are visible.}
\label{fig:tails}
\end{figure}

As a consequence, the tidal structures gather the matter that occupy a well-defined region in phase-space. Figure~\ref{fig:tails} (top row) shows the $N$-body toy-simulation of an encounter between a composite galaxy (disk+bulge+dark matter halo) and a point mass. Particles being part of one of the tails are tagged so that it is possible to track them back in time to their initial position in the disk. As mentioned above, these particles are distributed in a more or less confined region of phase-space at the time of the pericenter passage of the intruder, so that their individual motions are re-organized in a similar way. It is interesting to note that they cover a wide range of radii in the disk and thus, because of the differential rotation, the zone they occupy before the interaction is far from being symmetrical.

When the same experiment is repeated with an elliptical galaxy (Figure~\ref{fig:tails}, bottom row), the velocities are distributed almost isotropically and thus, no structure is created by the tidal field. As a conclusion, strong galactic tidal bridges and tails are formed from the material of disks galaxies. Note that the experiment we conducted above applies to any mass element, and thus can be, in principle, extended to both the gaseous and stellar components of a galaxy.

In the case of a flyby, the galaxies do not penetrate in the densest regions of their counterpart, do no loose enough orbital energy to become bound to each other, and thus they escape without merging. However, when the exchange of orbital angular momentum (through dynamical friction) is too high, the mean distance between the progenitors rapidly decreases (as a damped oscillation) before they finally merge, forming a unique massive galaxy. On the external regions of the merger, the tidal tails (if they exist) expand in the intergalactic medium and slowly dissolve. Because the tails are generally long-lived, they can indicate past interactions, as discussed in \citet{Struck1999}. As a result, tidal features can point to interacting events, even when what has caused their creation (i.e. a counterpart progenitor) has disappeared in a merger or has flown away.

\subsection{Gas dynamics}

The response of the gas to a galactic interaction can be seen as either an outflow or an inflow. For distant, non-violent encounters, a large fraction of the hot gas ($T > 10^3 \U{K}$) can be tidally ejected into the intergalactic medium, thus forming broad gaseous tails and/or halos around galaxies \citep[see e.g.][]{Kim2009}. It has been noted that while the least bound material would expand widely, more bound structures could easily fall back into the central region of the galaxies within less than $\sim 1 \Gyr$ \citep{Hibbard1995, Hibbard1996}.

During a first, distant passage, some galactic material is stripped off thanks to the transformation of the orbital energy of the progenitor galaxies. As a result, and because of dynamical friction, the interacting pair becomes more and more concentrated and can, under precise conditions \citep[see e.g.][]{White1978}, experience other passage(s) and finally end as a merger \citep{Barnes1992}. During such a second, closer interaction, tidal forces can induce shocks covering a large fraction of the galactic disk, which gives the gas a significantly different behavior than that of the stars \citep{Negroponte1983}. In particular, when stellar and gaseous bars form, the symmetry of the galaxy is broken: gravitational torques remove the angular momentum of this gaseous structure \citep{Combes1985} and make it fall onto the nucleus of the merger \citep[$< 1 \kpc$, see e.g.][]{Noguchi1988, Barnes1991, Hernquist1995, Mihos1996}. Such an inflow fuels the central region of the merger and participates in the nuclear starburst \citep{Springel2000, Barnes2002, Naab2006} often observed as an excess of infrared light or a strong nuclear activity \citep{Larson1978, Lonsdale1984, Soifer1984, Genzel2001, Younger2010}. At large radii in a disk, one gets the opposite effect: the gravitational torques push the material to the outer regions. This outflow enhances the formation of the tails already formed by the tidal field itself \citep{Bournaud2010}.

Note that star formation in mergers is also considered to be triggered by energy dissipation through shocks \citep{Barnes2004}. This is, however, quite sensible to the orbital parameters of the galaxies. Details about merger-induced starbursts are out of the scope of the present document. The reader can find a mine of information on this topic in \citet{Hopkins2006, Robertson2006, DiMatteo2007, Cox2008, Hopkins2009, Teyssier2010} and references therein.

Interestingly, \citet{Springel2005} showed that the collision between two gas-dominated disks could form a spiral-like galaxy instead of an elliptical one, as one could expect. In this case, a significant fraction of the gas is not consumed by the burst of star formation induced by the merger. Through conservation of the angular momentum, dissipation transforms the gaseous structure into a star-forming disk \citep{Hopkins2009}. Owning that the gas fraction in galaxy increases with redshift \citep[as suggested by][]{Faber2007, Lotz2010}, this last point sheds light on the formation history of low-redshift spiral galaxies.

\subsection{Influence of the internal and orbital parameters}
\label{sec:param}
The details on the formation of tidal structures are adjusted by several parameters that mainly concern the orbit of the galaxies, i.e. the way one sees the gravitational potential of the other. Because an analytical study of the influence of these parameters is very involved, many authors conducted numerical surveys to highlight the trends obtained from several morphologies.

\subsubsection{Spin-orbit coupling}

In their pioneer study, \citet{Toomre1972} already mentioned the influence of the spin-orbit coupling of the progenitors. For simplicity, let's consider two galaxies A and B separated by a distance $r_\mathrm{AB}$, and whose disks lie in the orbital plane. The norm of the velocity of an element of mass of the galaxy A situated at a radius $r$, relative to the galaxy B is $r_\mathrm{AB} \ \Omega \mp r \ \omega$, where $\Omega$ denotes orbital rotational velocity and $\omega$ the (internal) rotation speed of the galaxy A (i.e. the spin). The sign of the second term depends on the alignment of $\omega$ with $\Omega$. For a prograde encounter, the spin ($\omega$) and the orbital motion ($\Omega$) are coupled (i.e. aligned). Therefore, the relative velocity is lower ($r_\mathrm{AB} \ \Omega - r\ \omega$) than for a retrograde encounter ($r_\mathrm{AB} \ \Omega + r\ \omega$) and the net effect of the tides is seen for a longer period of time. As a result, the structures formed during prograde encounters are much more extended than those of retrograde passages.

Although this conclusion can be exported to inclined orbits, the strongest responses of the disks are seen for planar orbits, i.e. with a zero-inclination. The highly inclined configurations, called polar orbits, give generally birth to a single tail, as opposed to the bridge/tail pairs \citep{Howard1993}. In short, because an observed tidal effect does not only depend on the strength of the differential forces, but also on the duration of their existence, long tails are associated with prograde configurations. 

\subsubsection{Mass ratio}

Another key parameter is the mass ratio of the progenitors. In the hierarchical scenario, the galaxies form through the repeated accretion of small satellites \citep[see e.g.][and references therein]{Stewart2008}, and interactions between a main galaxy and number of smaller progenitors would occur more or less continuously. It is usual to distinguish the major mergers where the mass ratio is smaller than 3:1 (i.e. almost equal-mass galaxies), from the minor mergers involving a larger ratio (e.g. 10:1). In the last case, tidal tails are generally thin and small, while the same features are more extended and survive for a longer time in major mergers \citep{Namboodiri1985}.

The dependence of the structure of the remnant of the interaction (disky or boxy elliptical, as opposed to more symmetric galaxies) on the mass ratio of the progenitors has been extensively debated but is not directly connected to the tidal activity, and thus is out of the scope of this review \citep[see][for much more details]{Schweizer1982, Barnes1991, Barnes1992, Hernquist1992, Hernquist1993b,Naab2003,Bournaud2005,Bournaud2007}.

\subsubsection{Impact parameter}

During the interaction, the impact parameter plays an indirect role: a close, penetrating encounter will drive one galaxy deep inside the high density regions of the other, which implies a strong dynamical friction \citep[see e.g.][]{Bertin2003}. In this case, the separation of the progenitors after such a passage would be much smaller than for a more distant encounter.

Furthermore, a close passage generally corresponds to a significant tidal stripping. This situation occurs repeatedly for satellites orbiting within the halo of major galaxies \citep{Read2006}. Only the densest satellites can survive such a disruption \citep{Seguin1996}, while more fragile object would be converted into stellar streams \citep{Johnston1999, Mayer2002, Penarrubia2009}, as observed in the local Universe \citep{Ibata2001}.

However, the mass captured by a more massive companion (mass ratio close to 1:1) seems to be higher for short pericenter distances, as noted by \citet{Wallin1992}. The lost of material into the intergalactic medium is also higher under these circumstances.

\subsubsection{Dark matter halo}

In addition to the effect of orbital parameters, several authors noted the role played by the dark matter halo of the progenitor on the morphology of the merger, mainly the lenght of the tails. E.g. \citet{Dubinski1996} showed that long, massive tidal tails are associated with light halos, while the deep potential created by more massive ones would prevent the creation of extended structures. Note that, for a given mass, a dense halo appears to be more efficient in retaining the stellar component bound \citep{Mihos1998}. An important conclusion of this work was that galaxies exhibiting striking tails are likely to have relatively light halo (i.e. a dark to baryonic mass ratio smaller than $\sim 10:1$).

However, \citet{Springel1999} qualified this by stating that the important parameter is in fact the ratio of escape velocity to circular velocity of the disk, at about solar radius \citep[see also][]{Dubinski1999}. Therefore, even massive halos (e.g. mass ratio 40:1) can allow the growth of tails, provided the kinetic energy of the disk material is high enough to balance the depth of the gravitational potential of the massive dark matter halo. See Section~\ref{sec:DM} for more details.

\subsection{Rings, ripples, shells and warps}

Although they are the most visible structures formed during galactic interactions, the tidal tails and bridges are not the only signatures of encounters. Other mechanisms (not directly of tidal origin) lead to disrupted morphology. We briefly mentioned them here, for the sake of completeness.
\begin{itemize}
\item Shells or ``ripples'' describe the arcs and loops showing sharp edges in the envelope of galaxies. They originate from the collision between a massive galaxy and a small companion, 10 to 100 times lighter \citep{Quinn1984}. The material of the satellite is spread by an extensive tidal field in the potential well of the primary, along a given orbit of low-angular momentum \citep[see][and references therein]{Athanassoula1985}. A sharp ridge forms near the turnaround points of the orbit. The multiplicity of the shells corresponds to an initial spread in energy, leading to several possible radii for the ridges. 
\item A ring galaxy forms from the head-on collision between a large disk and a compact, small perturber \citep{Freeman1974, Theys1977}. The density wave created by the collision empties the central region of the disk and forms a ring in radial expansion (\citealt{Lynds1976}, see also \citealt{Appleton1996} for an observational and theoretical review).
\item Warped disks can be created by gravitational torques due to an infalling satellite galaxy \citep[mass ratio $\sim 10:1$, see][]{Huang1997,Revaz2001}. Note that warps and bending instabilities can also form through the torques exerted by a misaligned dark halo, or via accretion of matter \citep{Jiang1999, Bailin2003, Revaz2004}.
\end{itemize}

\subsection{Differences with tides at other scales}

The galactic tides are a purely gravitational effect, which means that they rely on scale-free quantities like the relative mass of the galaxies, the inclination of the orbits, their relative velocities and so on. Therefore, the conclusions presented above can be applied to any scales, from planetary to cosmological. If true in principle, this statement must be qualified because the requirements of the galactic-type tides themselves do not exist at all scales.

In the case of planetary tides, for example in the Earth-Moon system, the source of gravity does not penetrate in the object experiencing tides, and is generally situated at a distance large enough that it can be approximated by a point-mass. Furthermore, the binding energy of a solid and/or dense body like a planet is much higher than those of the galaxy on its stars. That is, the planetary tidal effects are weaker than the galactic ones. Note however that both the planetary and the galactic tides can destroy an object, like the comet Shoemaker-Levy 9 pulled apart by Jupiter's tidal field, or dwarf galaxies that dissolves in the halo of a larger galaxy, generally forming streams.

Another major difference arises from the periodicity of the motion. While a binary star or a planet is orbiting in a regular, periodic way, the galaxies show more complex trajectories, highly asymmetric, and rarely closed (because of high velocity dispersion and/or orbital decay). As a consequence, the tides at stellar or planetary scales can be seen as a continuous, or at least periodic effect, while they are rather well-defined in time and never occur twice the same way at galactic scales.

Therefore, the tidal effects seen at planetary or stellar scales, like the deformation of the oceans, atmospheres or external stellar envelops strongly differ from their equivalent phenomena in galaxies. At intermediate scale, the star clusters share properties of both tidal regimes. When orbiting an isolated galaxy, they undergo rather regular tidal effects and can, by filling their Roche lobe, evacuate stars through the Lagrange points. As a results, some globular clusters exhibit tidal tails, as seen in observations and reproduced by simulations \citep[see e.g.][and references therein]{Belokurov2006, Fellhauer2007, Kuepper2010}.

\section{Multi-wavelength observations of tidal tails}
\label{sec:wavelength}

Tidal tails have originally been discovered on deep photographic plates (see Section~\ref{sec:pec}) revealing  the optical light emitted by stars. This monochromatic, black and white,  view hides the variety of components and physical processes  hosted by collisional debris. Their multi-wavelength observation and analysis  were boosted  in the 90s \citep{Schombert90}. We present here-below an overview of the recent colorful  view of tidal tails.

The average {\it optical color} of tidal tails is consistent with the bulk of their stellar population being  older than the interaction, and originally born in the disk of the parent galaxies. 
Tidal tails however host bluer regions, whose light is dominated by OB-type stars. Given the life time of OB stars (less than 10 Myr) and the typical dynamical age of tidal tails (100 Myr), the young stellar component has been formed in-situ. These  giant star-forming complexes are usually compact and appear detached from the rest of the tails, explaining why they  were once believed to be ejected galaxies (see Section~\ref{sec:controv}).
Knots of star-formation are responsible for the bulk of the {\it ultraviolet emission} also emitted by tidal tails. 
Star formation is partially hidden by dust cocoons. Heated dust causes  the {\it infrared emission} of tidal tails.  The formation of stars requires the presence of gas. The main reservoir is atomic hydrogen, detected through the emission of the {\it radio 21cm} hyperfine line;  it hosts pockets of molecular gas  in which stars are born and that are detectable in the  {\it millimetric domain}, using  emission lines of molecules such as  carbon monoxide.

This section emphasizes the importance of multi-wavelength observations for studies of  the physical properties of tidal tails, especially those formed during major mergers.

\begin{figure}
\includegraphics[width=\textwidth]{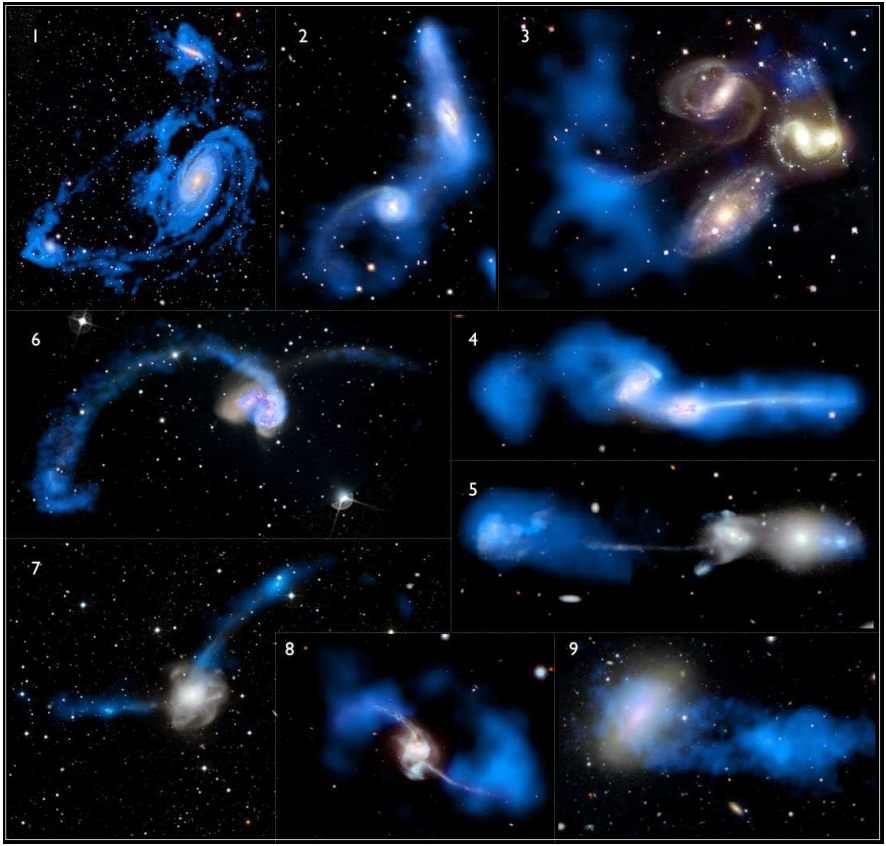}
\caption{A sample of interacting systems covering the various stages of major mergers, from the initial phases of the encounter (top left), to the last ones and formation of a relaxed object (bottom right). The gaseous component (atomic hydrogen) is superimposed on true color optical images of the galaxies, showing the distribution of the young and old stars. See Table~\ref{tab:sources} for details on the data.}
\label{fig:HI}
\end{figure}

\begin{table}
\caption{Observed systems and sources of data}
\begin{tabular}[h]{llp{7cm}}
ID & Name & Sources \\
\hline
1 & M81/M82/NGC~3077 & HI:VLA \cite{Yun94}; optical:DSS; \Ha: KPNO/36in \\
2 & NGC~2992/93 & HI:VLA \cite{Duc00}; optical:CTIO/SSRO (Courtesy D. Goldman); \Ha: ESO/NTT \\ 
3 & Stephan's Quintet & HI:VLA \cite{Williams02}; optical: HST (NASA); \Ha: Calar Alto/2.2m (courtesy J. Iglesias) \\
4 & The Mice (NGC~4676) & HI:VLA \cite{Hibbard1996}; optical: HST (NASA); \Ha: CFHT \\
5 & The Guitar (Arp~105) & HI:VLA \cite{Duc97b}; optical: CFHT; \Ha: CFHT \\
6 & The Antennae (NGC 4038/39) & HI:VLA \cite{Hibbard01}; optical: NOAO/AURA/NSF (B. Twardy); \Ha: CFHT + Palomar/1.5m \\
7 & The Atom for Peace & HI:VLA (Belles et al., 2012, in prep.); optical: ESO/WFC; \Ha: KPNO/2.1m\\
8 & NGC~2623 (Arp~243) & HI:VLA (Courtesy J. Hibbard); optical: HST/NASA/ESA (A. Evans); \Ha: CFHT \\
9 & NGC~4694 & HI:VLA \cite{Duc07b} (Courtesy VIVA collaboration); optical: ESO/NTT; \Ha: KPNO/0.9m \\
\hline
\end{tabular}
\label{tab:sources}
\end{table}

\subsection{Where the mass is: atomic hydrogen in tidal tails}
One of the first galaxies to have been fully mapped at 21~cm is the Antennae galaxies \citep[see also our Figure~\ref{fig:HI}, panel 6]{vanderHulst79}. These early observations obtained with the Westerbork Synthesis Radio Telescope (WSRT) revealed that about 70\% of the total amount of hydrogen in the galaxy pair is distributed along the optical tidal tails. For comparison, tidal tails account for only a few percent of the stellar component of colliding galaxies. 
The \hi\ gas appears as the principle, most massive, ingredient of tidal tails and is thus one of its best tracer.

Furthermore, since the \hi component is almost always more elongated than the stellar disk (by a factor 2-5, depending on the morphological type of the parent galaxy), it is less gravitationally bound than the stellar disk. As a consequence, gaseous tails are more easily produced than stellar ones. 
\cite{Hibbard95} used the Very Large Array (VLA) to carry out one of the first systematic study of \hi in pairs of galaxies. Observing systems of the so-called Toomre sequence (see  Figure~\ref{fig:toomre_sequence}), he was able to reconstruct the evolution of the gaseous component during a merger. Together with numerical simulations, these data show how part of the gas is stripped along the tails, while a fraction of it sinks into the central regions, sometimes via a bar and fuels there a nuclear starburst or an active galactic nucleus.
Finally, observations of the 21~cm \hi line have the additional advantage of providing the radial velocity over large scales. As emphasized in Sect.~\ref{sec:param}, a large variety of orbital parameters and corresponding models should be explored to reproduce the morphology of interacting systems. Having the complete radial velocity field restricts the parameter space. Tidal tails are too diffuse to allow spectroscopic measurements in the optical regime except in compact HII regions. Emission line regions are however not numerous enough in tidal tails  to allow a correct sampling of the velocity field, contrary to the \hi probe.

\subsection{When  components are missing: \hi without optical counterparts; stellar tails without gas}
Since the pioneer \hi observations mentioned above, numerous colliding systems have been mapped with the WSRT, the VLA, the Australia Telescope Compact Array (ATCA) or the Giant Metrewave Radio telescope (GMRT). In a vast majority of cases, there is a very good match between the \hi and the stellar components. The old stellar components and gas perfectly overlap, whereas young stars are formed at the \hi peaks. In a few rare cases, an offset is observed between the gas and the stars (see e.g. System~8 on Figure~\ref{fig:HI}). The origin of the star/gas  offset is debated: it may be due to different initial distributions of both components \citep[e.g.][]{Hibbard99} or additional processes that act on one component and not the other. For instance, ram pressure due to the interaction with the intergalactic medium might  strip the \hi gas further away \citep{Mihos01}.

Meanwhile, blind \hi surveys have disclosed the presence of numerous intergalactic, filamentary, \hi structures apparently devoid of stars \citep[see the review by][and references therein]{Briggs04}. 
Some may be of tidal origin. Spectacular examples are visible in the M81 group of galaxies (see System~1 in  Figure ~\ref{fig:HI}). Looking at the optical image of the M81 field, it might be difficult to infer that the three visible main galaxies are involved in a tidal interaction. The \hi map of the same region provides a different picture and reveals a complex network of tails and bridges connecting the three galaxies. 
The \hi Rogues gallery compiled by J. Hibbard\footnote{http://www.nrao.edu/astrores/HIrogues/} exhibits similar cases, emphasizing the role of \hi as the most sensitive tracer of on-going tidal interactions.

\begin{figure}
\includegraphics[width=\textwidth]{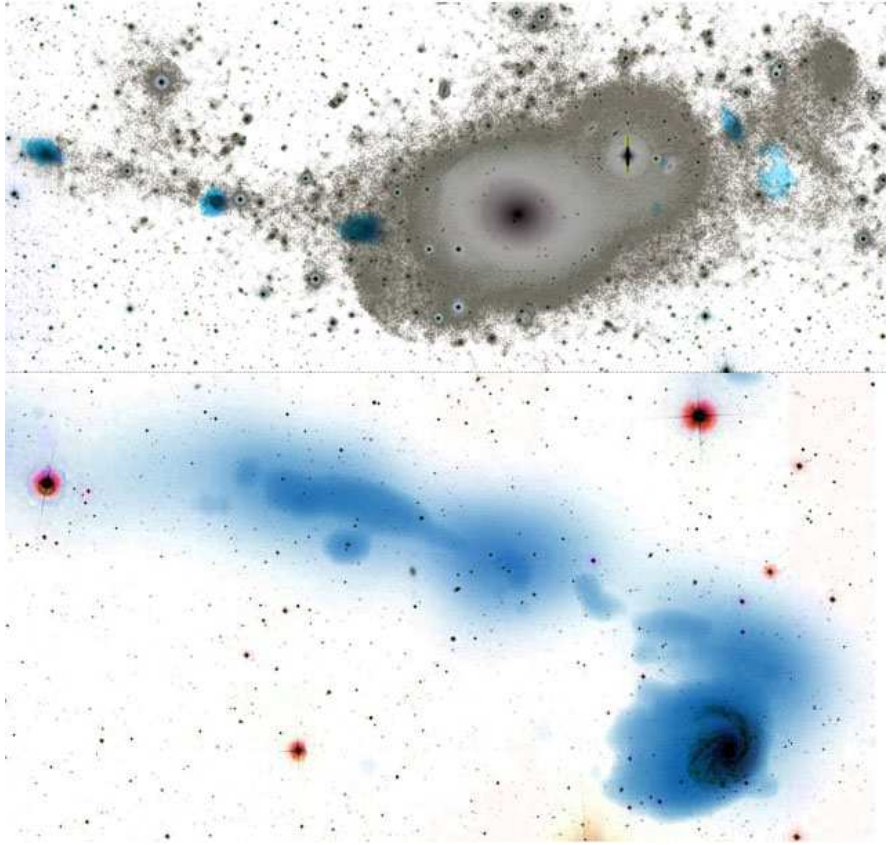}
\caption{Examples of a gas-poor and gas-rich tidal tails:  NGC~5557 \citep[top,][]{Duc11} and NGC 4254 \cite[bottom,][]{Duc08}, The \hi component is superimposed in blue on optical images. For both systems, the tails result from an encounter with a massive galaxy, which has merged in the case of  NGC~5557 or just flied by for  NGC~4254. The tidal tails of NGC~5557 are best visible in the optical as extremely low surface brightness structures, whereas the \hi emission is patchy and concentrated towards a few optical condensations.  The tail of NGC~4254 has no optical counterpart and was once believed to be part of a dark galaxy, known  as VirgoHI21. }
\label{fig:gas-poor}
\end{figure}

As a matter of fact, this may be a too  simple picture. Recently the optical regime had its revenge: with the availability of sensitive, large field of view CCD cameras, the surface brightness limit reached in the optical has gained several magnitudes. Diffuse light up to 29 mag.arcsec$^{-2}$ can be probed \citep{Mihos05, Ferrarese11}. In the nearest systems for which individual stellar counts are possible (with current technologies, in the Local Group), limits of 32 mag.arcsec$^{-2}$ are reachable \citep{McConnachie09}. At these limits, the most massive \hi tails do exhibit a stellar counterpart. This is likely the case for the M81 group \citep{Mouhcine09}, and many other interacting galaxies with available ultra-deep optical images \citep{Duc11}. An example of a newly discovered optical tidal tail, discovered as part of the Atlas$^{3D}$ survey    \citep{Cappellari11} is shown on  Figure~\ref{fig:gas-poor} (top pannel).

\begin{figure}
\includegraphics[width=\textwidth]{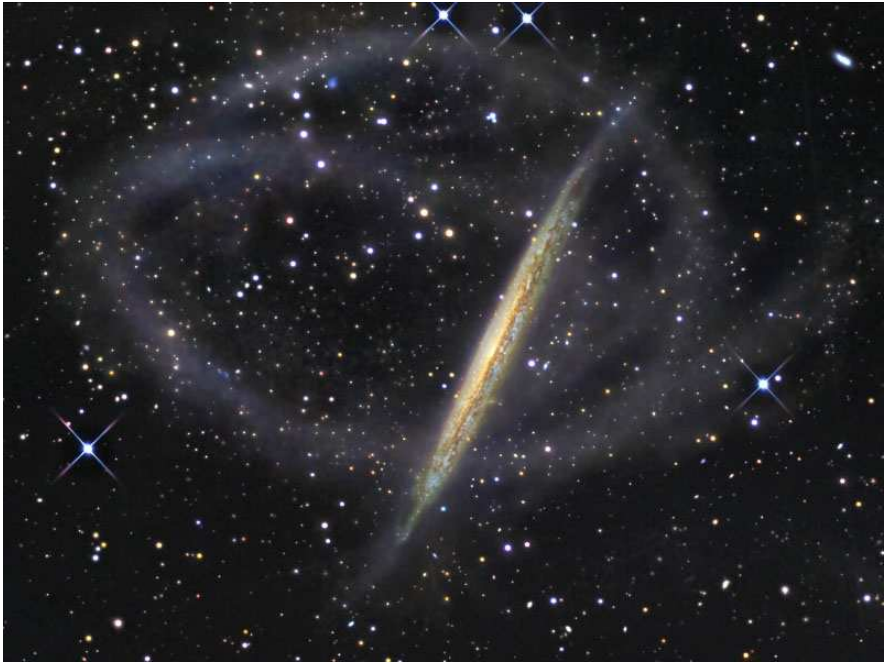}
\caption{Faint stellar streams wrapping around the spiral galaxy NGC~5907, seen edge-on. They are most probably due to a disrupted small dwarf spheroidal satellite. Such minor collisions are quite common around galaxies, including around our own Milky Way. Courtesy of   R. J. Gabany  in
collaboration with  \cite{Martinez-Delgado10}.}
\label{fig:stream}
\end{figure}

\hi intergalactic structures without any stars are thus much less frequent than once believed. A few of them however have escaped an optical detection. The Magellanic Stream in the local group is the most famous of them. This \hi structure is the largest tidal tail detected in the Local Group \citep{Nidever10}. It has for long been attributed to a tidal interaction between the Magellanic Clouds and possibly our Milky Way \citep[e.g.][]{Connors06}. However the absence of stars along the stream\footnote{Though, a stellar component has been found associated to the Magellanic Bridge.} was used to claim that this structure might in fact result from ram-pressure \citep{Moore94}. Indeed, ram-pressure only acts on the gas. The \hi is stripped along filaments that may be mistaken with tidal tails. Instances of long \hi tails, likely of ram pressure origin, may be found in the Virgo cluster \citep{Chung07}.
However, flybys, i.e. collisions  at high velocity which do not result in a merger, might as well produce tails without any stars, provided that the companion is massive enough to grasp gas from the target galaxy but resulting gravitational forces  too weak to drag the stars \citep{Duc08}. Such tails are of tidal origin but could themselves be mistaken with filaments created by ram-pressure.
According to a more exotic scenario, star-less intergalactic \hi clouds might reveal the presence of so-called ``dark galaxies", i.e. galaxies embedded in a massive dark matter halo that would contain very few baryons, only in the form of gas  \citep[see the proceedings of IAU symposium 244 dedicated to dark galaxies and][]{Minchin07}. An example of such objects is VirgoHI21, near the spiral NGC~4254 shown on Figure~\ref{fig:gas-poor} (bottom panel). The elongated cloud exhibits a strong velocity gradient, as if it was rotated and moved by an unseen dark component. But here again, such a velocity field might be explained by streaming motions generated by a tidal collision \citep{Bekki05,Duc08}. 

It is not unlikely that the isolated \hi clouds found in deep surveys such as the Arecibo Legacy Fast ALFA Survey \citep[ALFALFA,][]{Kent07} are simply collisional debris. 
Finally, another interpretation has recently gained popularity: the clouds and filaments around galaxies might divulge accretion of gas from so-called cold filaments. Simulations and some theoretical models emphasize the key role of external accretion of gas in the evolution of distant galaxies \citep{dekel09}. 
Primordial accreted clouds should be devoid of stars and have a low metallicity, whereas the metallicity of tidal debris should be high. This characteristics provides a method to disentangle tidal and cosmological origins for starless gas clouds. In practice the measure of element abundances  is extremely difficult for objects with no optical counterpart. \\

Stellar tails without any gas are rather common around massive galaxies. 
 Usually such streams are rather narrow and associated with tidally disrupted  satellites. The gas of the progenitors might have been stripped, evaporated or consumed well before the satellites were destroyed by their giant hosts. Stellar streams are regularly discovered in our own Milky Way: the Sagittarius and Monoceros streams are among the most famous ones \citep{Ibata01,Yanny03,Belokurov06}. Numerical simulations show how a satellite might be stripped of its stars, wrap around the main host galaxy before eventually falling in \citep[e.g.][]{Mayer01}.   A spectacular example of a disrupted dwarf in the halo of a spiral galaxy is shown on Figure~\ref{fig:stream}.

\subsection{Sparse components: molecular clouds, dust and heavy elements}
If old stars and \hi are the principle contributors to the mass of tidal tails,  they are accompanied by many other additional components. In fact collisional debris contain all the usual constituents of the interstellar medium of galaxies. A key ingredient is obviously the molecular gas in which stars are formed. \citet{Braine00} reported the first detection of carbon monoxyde at the tip of two tidal tails. Surveys of  colliding galaxies with HI-rich tidal tails lead to several other detections \citep{Smith01,Braine01,Lisenfeld02}. Follow-up CO(1-0) mapping with interferometers has been achieved in a few systems \citep{Lisenfeld04,Walter06}. CO clouds were detected towards local \hi peaks (and HII regions), with observed \HH/\hi mass ratio ranging from 0.02 to 0.5. This supports the hypothesis that the molecular gas has been formed locally out of collapsing \hi clouds \citep{Braine00}. However the later-on CO mapping of entire tidal tails revealed the presence of molecular clouds outside the \hi peaks, leaving open the possibility that the molecular component (or part of it) might have been directly stripped form the colliding galaxies at the same time as the \hi and stellar components \citep{Lisenfeld08,Duc07b}. 

The detection of CO at the tip of tidal tails indirectly reveals that heavy elements are present in that environment. The oxygen abundance could be determined in HII regions located along the tails \citep[e.g.][]{Weilbacher03}. Typical values are between one third and half solar, even at distances of 100 kpc from the parent galaxies. For comparison,   abundances in the very outskirts of isolated spiral galaxies range between one tenth to one third solar \citep{Ferguson98,Bresolin09}. The disk of spiral galaxies usually exhibit a strong metallicity gradient, with a possible flattening in the outmost regions \citep{Bresolin09};  no such gradient has yet been measured in tidal tails \citep{Kewley10}. 

The presence of cold dust in tidal tails has been first disclosed on far-IR images obtained with the ISO satellite \citep{Xu03}. Dust continuum emission in collisional debris has later-on been mapped by Spitzer \citep{Smith07,Boquien09} and more recently by Herschel. Furthermore the star-forming regions along tidal tails also exhibit mid-infrared emission features associated to polycyclic aromatic hydrocarbon (PAH) grains \citep{Higdon06,Boquien09}.\\

How did tidal tails acquire their metal-enriched components? Local stellar feedback during in-situ star formation episodes  contribute to the metal production.
However the onset of star-formation in collisional debris is likely too recent and the star-formation rates too small (see below)  to explain the  measured abundances in  heavy elements. Another hypothesis is a global enrichment of the interstellar/intergalactic medium  by stellar superwinds or enhanced AGN activity in the core  of the merging galaxies. Nuclear outflows might eject metal-enriched matter (in particular dust) up to large distances, as observed for instance in M82 (System~1 on Figure~\ref{fig:HI} and \ref{fig:SFR}). 
Alternatively, radial gas mixing during galactic collisions might account for the lack of metallicity gradients in tidal tails and the presence of dust at large galactocentric distances, as recently shown by numerical simulations of mergers \citep{Rupke10b}.

\section{Structure formation in tidal tails}

\subsection{Star formation}
As mentioned in the previous section, tails contain all the necessary ingredients for the onset of star-formation, in particular molecular gas and dust, and indeed young stars are often observed in collisional debris.

\begin{figure}
\includegraphics[width=\textwidth]{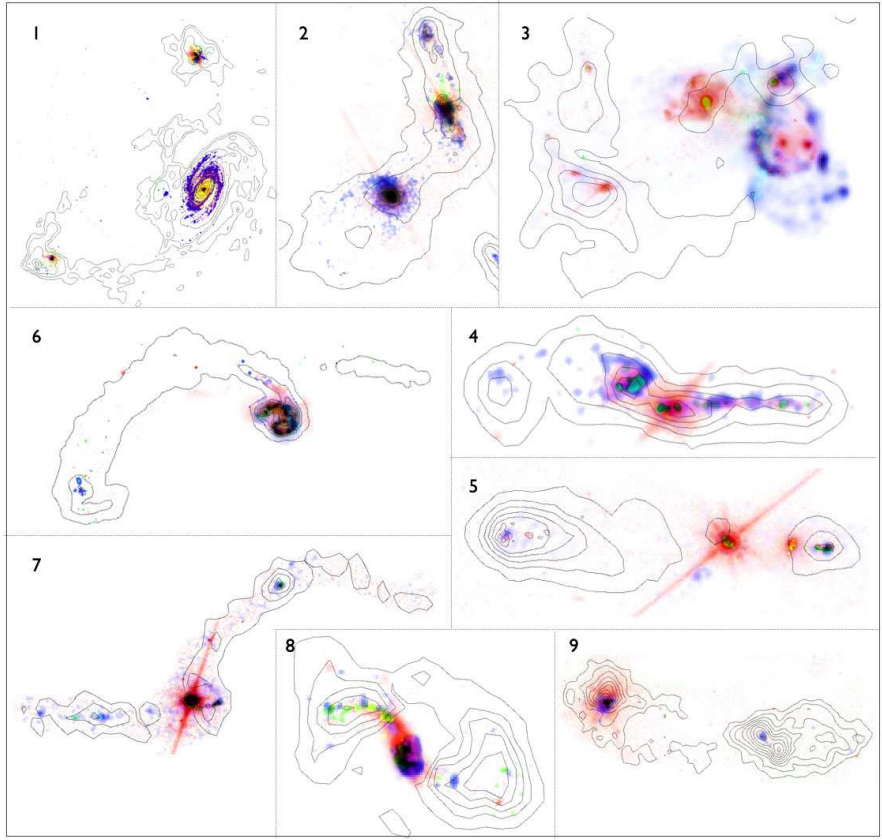}
\caption{Star formation in a sequence of merging galaxies. The displayed systems are the same as in  Figure~\ref{fig:HI}. The contours of the \hi 21cm emission are superimposed on a composite image combining light emission from three tracers of on-going or recent star-formation: the ultraviolet (blue), the H$\alpha$ line emission (green) and the mid-infrared (red). The most active star-forming regions belong to the so-called tidal dwarf galaxies.}
\label{fig:SFR}
\end{figure}

Census of star-forming regions in tidal tails has been carried out using a variety of tracers, such as the ultraviolet \citep{Boquien09,Smith10}, \Ha\ \citep{Bournaud04,Torres-Flores09} 
or mid-infrared emission \citep{Smith07, Boquien10}. These tracers may be combined  to further constrain the star formation history  (see composite image  on   Figure~\ref{fig:SFR}).
Star-forming regions in collisional debris may consist  of extremely compact and tiny knots with star formation rate (SFR) as low as 0.001 M$_\odot$/yr (see examples at the tip of the tails of  systems 3, 6 and 8 in Figure~\ref{fig:SFR})   or giant complexes with SFR reaching 0.1 M$_\odot$/yr (see  systems 5, 7 on Figure~\ref{fig:SFR}). 

A few studies have detailed the star-formation process in tidal tails, from the observational and theoretical   point of view \citep[e.g.][]{Elmegreen93}. Tidal objects are a priori  a special environment simultaneously characterized by (a)
the  same local chemical conditions as in spiral galaxies (ISM composition, metallicity)   (b)    the lack of an  underlying massive stellar disk, like dwarf irregular and low surface brightness galaxies  (c)  the kinematical conditions typical of mergers, i.e.  an enhanced gas turbulence and possibly shocks. 
Does then star-formation in collisional debris obey  the rules that prevail (a) in regular massive disks (b) in low-metallicity dwarfs, characterized by a low star-formation efficiency (SFE, the ratio between the star-formation rate and molecular gas content)  (c) in the central regions of mergers  where deviations from the so--called Kennicut-Schmidt relation (a correlation between the star-formation rate per unit area and the gas surface density) have been  measured 
 \citep{daddi10}?  The SFE estimated in several tidal objects favors the first hypothesis: its value is close to that usually measured in galactic disks  \citep{Braine01,Boquien11}.

Therefore, with respect to star-formation, tidal tails do not appear as exotic objects. The properties of the pre-enriched  interstellar medium inherited  from their parent galaxies  govern their star-formation capabilities rather than  the  violent episode at their origin or the large-scale (intergalactic) environment in which they now evolve.

\subsection{Star cluster formation}
\begin{figure}
\includegraphics[width=\textwidth]{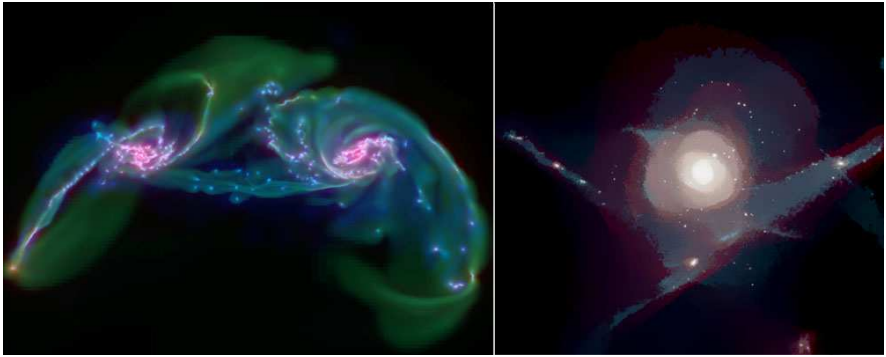}
\caption{Formation of stellar structures in high-resolution numerical simulations of major mergers. {\it Left}: after the first pericentric passage, with the hydrodynamical AMR code RAMSES  \citep{Teyssier2010}. {\it Right}: at the merger stage, with a sticky-particle code \citep{Bournaud08}. On the electronic version of this figure, gas is rendered in green, young stars in blue and old stars in magenta/brown. Both models show the formation of stellar objects (rendered in yellow/white): compact knots with properties similar as Super Star Clusters, or more massive, extended   structures resembling  Tidal Dwarf Galaxies.}
\label{fig:sim-sscs}
\end{figure}

Galaxy mergers do not only enhance star-formation. 
The increase of the gas pressure during mergers triggers the formation of star clusters as well. 
The Hubble Space Telescope has revealed the presence of a large population of young Super Star Clusters (SSCs) in nearby merging systems, including along tidal tails
\citep[see][for a review]{Schweizer06}.
The most massive of them are believed to evolve into globular clusters (GCs), thus making mergers a possible origin of GCs. 
Numerical simulations at high resolution support this hypothesis  \citep{Bournaud08}.  Figure~\ref{fig:sim-sscs} presents two different models that were able to form SSCs.  
Globally, the cluster formation rate follows the star-formation rate. The infant mortality of SSCs less than 10 Myr after their formation appears however to be very high. SSCs in particular suffer from sudden gas loss due to feedback effects that alter their dynamical stability. There are special locations in merging systems, where local compressive tidal modes might contribute to (at least partially) protect them and increase their life-time \citep{Renaud2009}. Large volumes (up to 10 kpc wide) of compressive modes have been located in the tidal tails of major mergers, with an intensity comparable to that found in the central regions. But the lower gas density and turbulence in such an environment do not seem to particularly favor the formation of SSCs in tails.

\subsection{Formation of Tidal Dwarf Galaxies}
Tidal tails host the most massive structures that may be born during galaxy mergers: the Tidal Dwarf Galaxies (TDGs). 
As indicated by their name, TDGs have the  mass of classical dwarf galaxies, i.e. above  $10^8~\Mo$. They have originally been detected on optical images as prominent and generally blue (thus star--forming) condensations at the end of tidal tails. Follow-up radio observations revealed that they were associated with massive \hi clouds \cite[see][for a recent review on TDGs]{Duc11b}.
Detailed kinematical studies of the ionized,  \hi or molecular gas indicate that TDGs are gravitational bound entities that are kinematically decoupled from their parent galaxies. They exhibit velocity curves that are typical of rotating objects.
In practice, the kinematical study of tidal tails suffers from strong projection effects: tidal tails are highly curved filaments; when seen edge-on, several components of the tail may be projected along the same line of sight. This creates an artificial velocity gradient that may be mistaken with a genuine rotation curve. Projection effects are especially critical near the end of tidal tails where most TDGs are precisely located \citep{Bournaud04}.

Numerical simulations have provided clues on the formation mechanism of tidal dwarf galaxies (see examples on Figure~\ref{fig:sim-sscs}). Several scenarios have been proposed: 
\begin{itemize}
\item growth of condensations born following local gravitational instabilities in the stellar component \citep{Barnes92} or in the gaseous component \citep{Wetzstein2007}\footnote{\cite{Wetzstein2007} claimed that the clumps formed in N-body models that do not include gas are numerical artifacts},
\item multiple mergers of super star clusters \citep{Fellhauer02},
\item formation and survival of massive star clusters thanks to the fully compressive mode of tidal forces \citep[][see above]{Renaud2009},
\item formation of massive gas clouds in the outskirts of colliding disks, following the increased gas turbulence, that become Jeans-unstable and collapse once in the intergalactic medium \citep{Elmegreen93},
\item accumulation and collapse of massive gaseous condensations at the end of the tidal tails, following a top-down scenario  \citep{Duc04b}.
\end{itemize}

\begin{figure}
\centerline{ \includegraphics[width=9cm]{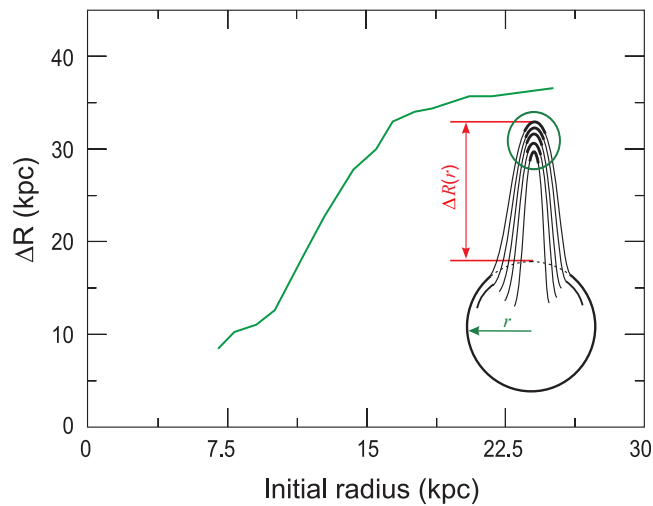}}
\caption{The effect of tidal forces on the potential well corresponding to an extended dark matter halo. Amplitude of the radial excursions of matter as a function of the initial radius in a numerical model made of concentric annuli. Above a certain distance, it becomes constant, enabling an accumulation of gas in tidal tails, the seed of tidal dwarf galaxies. Adapted from \cite{Duc04b}.} 
\label{fig:TDGform}
\end{figure}

In the context of this Review, we detail here the latter scenario as it grants to the shape of tidal forces a key role in the formation of TDGs.
In the potential well of disk galaxies, constrained by extended massive dark matter halo (see Section~\ref{sec:DM}), the tidal field carries away the outer material, while keeping its high column density -- the radial excursions are constant, as illustrated in  Figure~\ref{fig:TDGform}. Gas may pile up at the tip of tidal tails before self-gravity takes over and the clouds fragment and collapse. Toy models show that the local shape of the tidal field plays the key role in structuring tidal tails and enabling the formation of TDGs. 

The presentation of the long term evolution and survival of TDGs is behind the scope of this Review. Details on the predictions of numerical simulations and observations of old TDGs may be found in \cite{Duc11b}.

\section{Tidal structures as probes of galaxy evolution}
Tidal tails, and more generally the fine structures that surround galaxies (stellar streams, rings, bridges, shells) are among the least ambiguous signposts of galaxy evolution. Indeed, whereas other galactic properties such as the presence of spiral structures, bars, warps or even starbursts, may be accounted for by secular and internal evolution, the formation of stellar filaments can only be explained by a past collision between galaxies. Numerical cosmological simulations predict the formation of 
many such structures \citep[see among many others][]{Johnston08,Peirani10}. 
However, their census and interpretation face a number of issues.
\begin{itemize}
\item Fine structures are faint, with typically optical surface brightness fainter than 26 \sbr, HI gas column densities below $10^{19} $\cmm, and thus difficult to observe. Nevertheless current generation of optical surveys as well as deep blind \hi surveys have now the required sensitivity to detect a significant fraction of the large number of fine-strucures predicted by numerical models of galaxy evolution.
\item The properties of the fine structures depend on the properties of the parent galaxies: a wet merger (a collision involving gas-rich galaxies) will generate gaseous streams; stars from hot stellar systems (early-type galaxies) will not make long tidal tails. Prograde encounters produce more narrow tidal tails. Conversely, by studying the shape and inner characteristics of collisional debris, one may learn about the properties of their ancestors.
\item Fine structures may be short lived. It takes a few 100 Myr to form a long tidal tail and a similar time to destroy them: tidal material may be dispersed or fall back at a rate  that depends on the distance to the parent galaxy, from a few hundred Myr \citep{Conselice09} up to a few Gyr \citep{Hibbard1995}. Conversely the discovery of a tidal tail around an object might provide an age estimate of the last major merger event. With the support of a numerical model of the collision, one may even reconstruct the history of the collision (or at least have a model consistent with it) and predict its future.
\item Fine structures are fragile and quickly react to their environment. For instance, in clusters of galaxies, tidal tails appear more diffuse as the interaction with the additional potential well of the cluster will accelerate the evaporation of their stars. As a consequence tidal tails should be less visible in dense environments and larger stellar halos are expected there, which is indeed observed \citep{Mihos05}.
\end{itemize}

Such issues might be addressed by combining predictions from numerical simulations and observations. 
 We present below examples of the use of tidal structures as probes of galaxy evolution and the mass assembly of galaxies. 

\subsection{Determining the merger rate evolution with tidal tails}
\begin{figure}
\includegraphics[width=\textwidth]{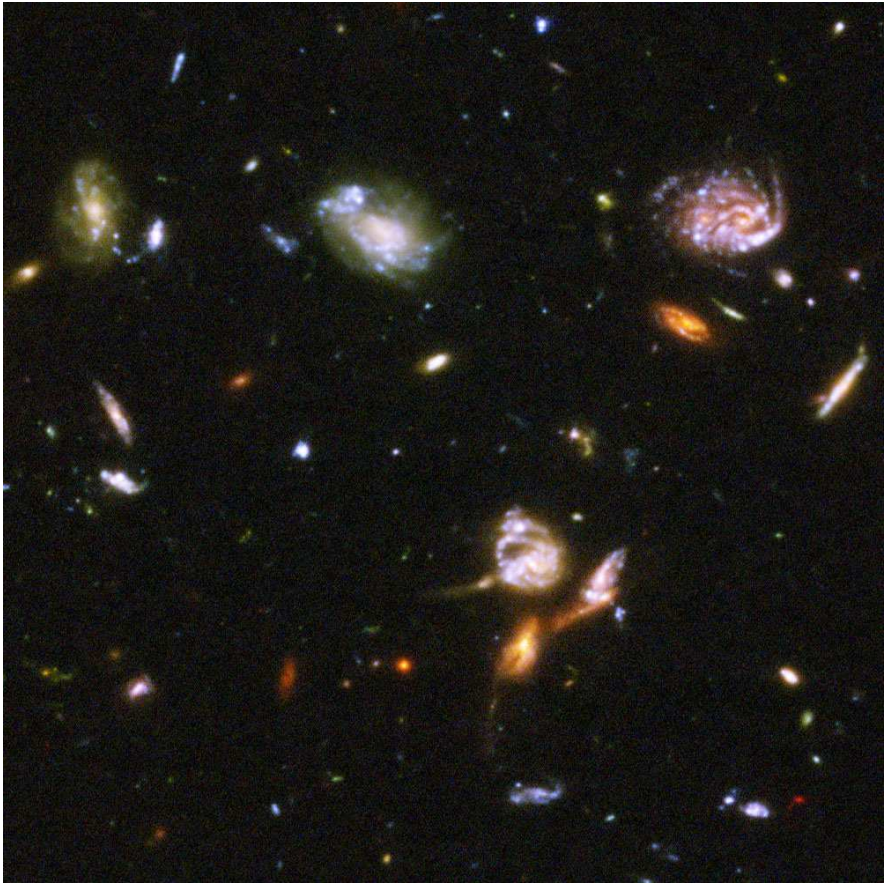}
\caption{The Hubble Ultra Deep Field, showing several tidally perturbed distant galaxies. The evolution of the fraction of collisions as a function of redshift is the subject of strong debates. Credit: NASA/ESA.}
\label{fig:UDF}
\end{figure}

Early deep observations with the Hubble Space Telescope revealed that distant galaxies ($z>1$) seemed to be much more morphologically perturbed than local galaxies \citep{Griffiths94,Glazebrook95,Abraham96b}, supporting the idea that a smaller, denser and younger Universe favored galaxy-galaxy collisions. Since then, many studies based on deep surveys, such as the one illustrated in Figure~\ref{fig:UDF}, have tried to quantify the evolution of the merger rate as a function of time, without in fact reaching a consensual value. A variety of methods have been used, based on:
\begin{itemize}
\item the census of close galaxy pairs \citep[e.g.][]{LeFevre00,Kartaltepe07}. The method assumes that galaxies observed in pairs are physically linked and doomed to merge.
\item the identification of perturbed kinematics using Integral Field Spectroscopy. The method has been recently used as part of the IMAGES \citep{Yang08}, MASSIV \citep{Contini11}, SINS \citep{Shapiro08} and AMAZE/LSD \citep{Gnerucci11} surveys at redshifts of 0.6, 1.3, 2 and 3 respectively. This method is very time consuming and may only be applied to limited samples. 
\item the census of morphologically perturbed galaxies showing for instance anisotropies in their stellar distribution \citep[e.g.][]{Conselice03b}. This requires a reliable algorithm to automatically measure the degree of perturbation. 
\item the direct detection of tidal tails \citep[e.g.][]{Bridge10}, which, as argued earlier, is likely the most direct technique. 
\end{itemize}
However a few remarks need to be made at this stage: first, the most massive component of tidal tails formed in major mergers is by far the atomic hydrogen. As mentioned before, \hi surveys might disclose collisional debris that are hardly visible in the optical. Unfortunately, the current technology and antennas sensitivity limit the detection of the  21~cm emission  line to redshifts less than 0.3. In the more distant Universe, tidal tails may only be observed through the emission of their stars. Intrinsic dimming with redshift as well as band shifting make them less and less visible and bias surveys in favor of  UV emitting, star forming structures. 
Other difficulties arise at high redshift.  The gas fraction of  galaxies was higher and their gaseous disks more unstable. Prominent  star forming condensations formed in the disk   may be mistaken with either multiple nuclei of merging galaxies or even condensations within tidal tails \citep{Elmegreen09}. Among these ``clumpy" galaxies, 
only a fraction of them (for instance the so-called ``tadpoles" systems) may be genuine interacting systems \citep{Elmegreen07b}.
One usual hypothesis when counting the number of tidally perturbed systems is that disk-disk collisions at low and high redshift produce similar external structures. However if the  colliding progenitors are the gas-rich clumpy disks mentioned above, the mutual interaction between their clumps (which have masses comparable to that of dwarf galaxies) might prevent the formation of tidal tails \citep{Bournaud11}.
Thus, when trying to measure the evolution of the past merger rate by looking at the level of tidal perturbations, one should keep in mind that distant tidally interacting galaxies might differ from those observed in the Local Universe. 

A last word of caution: when comparing the merger rate at low and high redshift, it is assumed that the fraction of galaxies involved in a tidal interaction is well known in the nearby Universe \cite[and considered not to exceed a few percent, see][]{Miskolczi11}. However even tidal tails from past major mergers might have been missed at $z=0$  because of their low surface brightness. 
Indeed, the extremely deep mapping of the Andromeda region, in the Local Group,  has revealed  an extremely faint  stellar bridge between M31  and M33 \citep{McConnachie09}, suggesting  that the two spirals are involved in a tidal collision. Prominent  tidal tails of very low-surface brightness  were also recently  discovered around apparently relaxed massive ellipticals  \citep{Duc11}. An example is shown on the top panel of  Figure~\ref{fig:gas-poor}. 
Such observations indicate that, in the local Universe, the fraction of tidally interacting galaxies is likely underestimated: 
serious issues plague the determination of the merger rate even at low  redshift. 

\subsection{Determining the mass assembly history of galaxies with tidal tails}
The existence of a tidal tail unambiguously establishes the occurrence of merger event in the past history of the host galaxy. Therefore, the census of the collisional debris might, in principle, constrain the recent mass assembly of nearby galaxies. Now that surface brightness limits of unprecedented depth may be reached with the current generation of optical cameras with large-field of view, this method of galactic archeology might be very powerful. However, it faces a number of issues:
\begin{itemize}
\item The frequency of tidal features, and degree of tidal perturbation (which one would like to link to other properties of the parent galaxies to constrain their mode of formation), is difficult to quantify. Tidal indexes proportional to the degree of morphological asymmetries have been introduced \citep[e.g.][]{Tal09}; most often however, more subjective ``fine structure'' indexes determined by eye are used to classify merging pairs or more evolved systems \citep{Schweizer90, Schweizer92b}.
 \item Not all collisions and mass accretion events produce tidal tails. The method is biased against mergers involving hot, pressure supported, galaxies. Indeed tidal forces most efficiently act on (rotating) disks. This indirectly means that tidal tails trace wet mergers rather than dry ones.
\item Tidal features fade with time, either because they fall back onto their progenitors or evaporate into the intergalactic medium. Their detectability, and the ability to trace back past merging events, strongly depend on the surface brightness limit achieved by the observations.
\item The destruction rate of tidal features depends on the environment. Dense environments such as galaxy clusters contribute to erase collisional debris \citep{Tal09}. Tidal tails may also be destroyed during successive merger events. 
\end{itemize}
As a consequence, it might be difficult to probe collisions older than a few Gyrs.

\subsection{Constraining the distribution of dark matter with tidal tails}
\label{sec:DM}
Tidal tails might not only tell us about the baryonic content of their parent galaxies and how it reacted to the environment; they are as well insightful to constrain the structure and distribution of the most massive component of galaxies: dark matter (DM). Rotation curves of galaxies reveal how much gravitational matter is located within the radius at which velocities are measured but do not constrain the extent and 3D shape of the dark matter halo. The halo of  CDM models is very extended, at least 10 times the optical radius. The rotation curve cannot be easily probed at these large  distances. Tidal tails produced during major mergers have however sizes that can exceed 100 kpc, reaching the outskirts  of the dark matter halos: tails are thus a priori a convenient tool to probe the structure of cosmological  halos. Numerical simulations have been used to study the effect of the size of the DM halo on the shape of tidal tails. Apparently contradictory results have been obtained, claiming or not a dependence with the halo mass, size, concentration or spin \citep{Dubinski1996,Mihos1998,Dubinski1999,Springel1999}.

The shape of the DM halo, its triaxiality and presence of sub-halos might be probed by smaller, thiner tidal tails from minor mergers that wrap around galaxies. Those found around the Milky Way, such as the Sagittarius stream, are the target of numerous studies \citep[e.g.][]{Mayer2002,Helmi04,Penarrubia06,Varghese11}. 

While no direct correlation between the size of the DM halo and the size of tidal tails has yet been established, the internal structure of tidal tails might be connected to the DM extent. \cite{Bournaud03} argued that the massive condensations at the tip of tidal tails, associated with TDGs, cannot be formed if the halo of the parent galaxy is truncated. \citet{Duc04b} provided a toy model showing that in the  case of a truncated halo, the tidal material is stretched along the tidal tails, preventing its collapse and the formation of massive sub-structures. When the halo is large enough, this stretching does no longer occur beyond a certain distance, and apparent massive condensations near the tip of the tail might form TDGs (see Figure~\ref{fig:TDGform}).
The observation of TDGs is thus consistent with the extended dark matter halos predicted by the CDM theory.

If large DM halos seem to be required to form TDGs and shape the inner structures of tidal tails, tails should themselves  not contain large quantities of dark matter. Indeed the current picture of DM makes them collisionless particles distributed in a hot halo on which tidal forces have little impact. The tidal material originates from the disk, which is predicted to contain almost no DM.  In practice, the DM content of tidal tails is difficult to probe. However in some special circumstances, it may be measured using the traditional method of rotation curves. Tidal dwarfs are gravitationally bound systems; their DM content may thus simply be derived determining their  dynamical mass and subtracting it from the luminous one (consisting of HI, \HH, stars and dust). This exercise has been carried out for a few systems \citep[][Belles et al 2012, in prep.]{Bournaud07,Duc07b}. Even if the error bars are large, these measurements  yield reliable dynamical to luminous mass ratios of 2--3.
Assuming that the CDM theory is correct, one should conclude that TDGs (and thus the galactic disks) contain non-conventional dark matter, likely traditional baryonic matter  which has not yet been detected by existing surveys.  A possible candidate  is  molecular gas not accounted for by CO observations. The observations of dust in the far infrared by the Planck satellite  supports the hypothesis of an unseen, dark, component in the gaseous disk of galaxies, which might contribute to the global budget of the missing baryons in the Local Universe \citep{Planck11}.
Alternatively, CDM might be wrong, as claimed by several groups who push for modified gravity. Modified Newtonian Dynamics (MOND) has retrieved the rotation curves of galaxies, including TDGs, without the need of a dark matter halo \citep{Milgrom07,Gentile07}. Numerical simulations of galactic collisions in the MOND framework have been carried out: they also reproduce the long tidal tails made with classical Newtonian dynamics \citep{Tiret07}. The main difference is the absence of dynamical friction during the collision, which contributes to extend the time scale of the collision, and decrease the probability of a final coalescence.

\section{Conclusions}
It is an undeniable fact that tidal forces and the formation of tidal tails are overall a second order  process in galaxy evolution. The fraction of stars expelled in the intergalactic medium is low, at most a few percent in major mergers. The fraction of gas is more important, but the bulk of the gaseous reservoir is funneled into the central regions. Collisional debris may host star-forming regions, but their contribution to the total star formation rate is minimum. Clearly, most of the activity occurs in the more central and nuclear regions where starbursts and/or AGN fueling is triggered. 
However, one of the aims of the present Review is to emphasize the idea that tidal debris can provide insightful information about the properties of galaxies, the same way as garbage in trash cans tells us much about the way of life of their owners.

The presence of tidal features is an unambiguous proof that a major/minor merger occurred in the recent past, and that at least one of the colliding galaxies had a stellar and/or gaseous disk. The converse  is not true though,  as not all collisions produce prominent tidal features. Determining when the merger took place is less strait-forward. However, numerical simulations done in  cosmological context will soon be able to constrain the survival time of collisional debris and thus give predictions on their age. Comparisons between observations and simulations should then allow us to reconstruct the mass assembly of galaxies.  Current generation of wide-field-of-view cameras and the on-going extremely deep surveys of the nearby Universe detect numerous new tidal features of very low surface brightness, offering interesting prospects to galactic archeology. At high redshift, the census of tidal perturbations is much more complex, not only because of dimming and band shifting issues, but also because distant galaxies are much more gas--rich and therefore are intrinsically irregular. This makes the separation between secular and external effects rather ambiguous.
 
Multi-wavelength surveys  have revealed the presence in collisional debris  of all the constituents of regular galaxies  though with different proportions: young and old stars, atomic gas, molecular gas, even possibly dark gas, heavy elements and dust. Star formation seems to proceed there in a similar way as in isolated spiral disks, despite the very different environment at large scale.
 
Tidal tails may in principle even be used to probe some fundamental aspects of physics, including, of course, the properties of tidal forces but also the laws of gravitation, as shown by recent experiments with modified gravity. 
The fact that tidal forces can be compressive and for instance  contribute to the stability of star clusters whereas they are usually associated with destruction processes has only recently been understood. The shape of the tidal tensor explains why massive tidal dwarf galaxies may only be formed within an extended dark matter halo. A theoretical study on the nature and the role of tidal forces in galaxies remains largely to be done and might provide further surprises. 

\section*{Acknowledgments}

First of all, we express our gratitude to the two main organizers of the school, Jean Souchay and St\'ephane Mathis. We not only enjoyed the marvelous premises --  Cargese in Corsica --, but also the very stimulating discussions that took place between experts in terrestrial, planetary, stellar and galactic tides. We are very grateful to Fr\'ed\'eric Bournaud for daily discussions on galaxy collisions and numerical simulations. We finally wish to thank all our collaborators and colleagues for their crucial contributions to the various works presented here. 

\bibliography{bib_florent,bib_pa}

\end{document}